\documentclass[%
 reprint,
 amsmath,amssymb,
 aps,
floatfix,
]{revtex4-2}

\usepackage{graphicx}
\usepackage{dcolumn}
\usepackage{bm}
\usepackage[x11names]{xcolor}
\usepackage[percent]{overpic}

\usepackage{pict2e}

\usepackage{siunitx}
\DeclareSIUnit\molar{\text{M}}

\usepackage[normalem]{ulem}

\begin{document}


\newcommand{\conc}[1]{[\mathrm{#1}]}
\newcommand{\Kd}[1]{K_{\mathrm{D}}^{\mathrm{#1}}}
\newcommand{\Ka}[1]{K_{\mathrm{a}}^{\mathrm{#1}}}
\newcommand{\DF}{\mathrm{DF}}
\newcommand{\koff}[1]{k_{\mathrm{off}}^{\mathrm{#1}}}
\newcommand{\kon}[1]{k_{\mathrm{on}}^{\mathrm{#1}}}
\newcommand{\kb}{k_{\mathrm{B}}}
\newcommand{\kT}{k_{\mathrm{B}}T}
\newcommand{\cc}[1]{c_{\text{#1}}}
\newcommand{\ts}{\textsuperscript}



\title{Membrane-Associated Self-Assembly for Cellular Decision Making}
\author{Samuel L. Foley}
  \email{sfoley13@jhu.edu}
  \affiliation{T. C. Jenkins Department of Biophysics, Johns Hopkins University, Baltimore, Maryland, USA}
\author{Margaret E. Johnson}%
  \email{margaret.johnson@jhu.edu}
  \affiliation{T. C. Jenkins Department of Biophysics, Johns Hopkins University, Baltimore, Maryland, USA
}

\date{\today}

\begin{abstract}
Cellular decision-making based on information received from the external environment is frequently initiated by transmembrane receptors. These receptors are known to propagate such information by triggering a series of irreversible, energy-consuming reactions. While this active mechanism ensures switch-like responses, here we show how spontaneous self-assembly of native 3D subunits on a two-dimensional substrate can similarly act as a tunable and robust switch for detecting receptors at physiological concentrations. This mechanism is much more sensitive than other passive mechanisms for receptor detection. We derive analytical expressions for the critical receptor density driving stable subunit assembly, in close agreement with stochastic reaction-diffusion simulations. The theory provides testable predictions for how lipids, subunits, and receptors each can control decision boundaries and magnitude of response.

\end{abstract}

\maketitle

\section{Introduction}

Living systems receive diverse inputs from their surrounding environment, including neighboring cells, via a variety of molecules.
These molecules, including essential nutrients\cite{barata2021endocytosis}, growth factors\cite{sorkin1993endocytosis}, and surface adhesion proteins\cite{cadwell2016cadherin}, interact directly with transmembrane receptors bridging the outside of the cell to the cell interior\cite{lauffenburger1996receptors}. To initiate the appropriate output response, the familiar form of information processing is for these local interactions to initiate a series of irreversible reactions triggered by enzymes, generating a selective and switch-like transition\cite{ferrell2014ultrasensitivity, tkavcik2016information, franccois2016case, tkavcik2025information}. However, given the diversity of inputs and subsequent decisions (e.g. nutrient uptake), can alternate molecular paradigms for receptor detection at a threshold density decouple sensing from energy expenditure while retaining the switch-like response? Endocytosis\cite{mettlen2018regulation,mondal2022multivalent}, adhesion formation\cite{malinverno2017endocytic,troyanovsky2023adherens}, and immune cell-antigen binding\cite{mcaffee2022discrete} may not be classically considered ``information processing'', but each require detection of receptors at controlled times and places for proper function. Because each of these processes combine receptor binding with self-assembly of cytosolic proteins, we propose that the assembly itself offers a reliable way to encode a decisive, localized response. We show here that equilibrium binding of subunits to receptors, which does not induce a sharp sensitivity to density changes, can be pushed to a switch-like response via self-assembly of subunits benefiting from a 2D search space.

The equilibrium self-assembly of subunits with a sufficient valency $Z>1$ and topology can transition sharply from free to assembled over small changes in concentration or binding free energy  $\Delta G$. This stems from reduction in the probability of subunit dissociation by $\exp(\Delta G/\kT)$, where $\kT$ is the thermal energy, following the formation of a closed loop of subunits\cite{jhaveri2024discovering,ercolani2003assessment, zlotnick1994build}. This cooperativity is emergent rather than explicitly built-in through allostery, which we exclude from the assembly\cite{jacobs2015rational,jacobs2021self,zlotnick1994build,zhong2017associative,evans2024pattern}. We exploit this sensitivity to concentration changes by coupling lattice assembly to dimensional reduction on the surface\cite{adam1968reduction} and subsequent receptor binding. Receptor detection thresholds thus emerge in our theory and simulation from reversible bimolecular interactions via the participation of discrete, multi-valent subunits, without the chemical feedback common in pattern formation\cite{wurthner2022bridging}. The role of the receptor is related to, but distinguishable from, membrane adsorption. By providing a second anchor to the surface following adsorption, receptors increase 2D subunit densities, promoting assembly that feeds back into reduced desorption via closed loops with the surface; hence there is an accumulation of effective cooperativity. Our framework makes the dependence on the volume-to-surface-area ratio $\ell \equiv \mathcal{V}/\mathcal{A}$ transparent, as together with the changes to $\Delta G$ from 3D to 2D\cite{xu2015binding}, these factors set the maximal gain in assembly yield from concentrating subunits in 2D\cite{yogurtcu2018cytosolic}. Patterns and nucleation observed in models\cite{wurthner2022bridging,guo2022large} and experiments\cite{toprakcioglu2022adsorption, brauns2021bulk} are strongly coupled to $\ell$.
    
Achieving assembly-driven sensing and discrimination of receptor densities without activation or energy input requires proper tuning of interaction strengths relative to concentrations. Subunits should be native 3D species with concentrations below the critical threshold for nucleation. Adsorption and confinement to the membrane\cite{araujo2023steering} will move them towards the critical concentration\cite{yogurtcu2018cytosolic}, but to be effective sensors, they should remain below this concentration in the absence of receptors. We derive a free energy function that predicts the fraction of subunits in ordered lattices for a finite pool of subunits. Without receptors, our model resembles pre-wetting frameworks\cite{rouches2021surface,zhao2021thermodynamics,kaplan2013review}, albeit with the dense (lattice) phase restricted to 2D topologically. A primary difference (in addition to finite size) is that our free energy derives from specific discrete pairwise interactions, such that our parameters then map directly to experimental measureables like binding affinities; the continuum nature of pre-wetting\cite{cahn1977critical,goldenfeld2018lectures} renders physiologic relevance of phase-separating regimes hard to test. 
We derive an accurate closed-form approximation to the critical receptor threshold from our theory and quantitatively test it via stochastic simulations. Our results show that receptor sensing by 2D self-assembly is broadly achievable in living systems at realistic timescales.


\begin{figure}
    \centering
    \begin{overpic}[width=0.98\linewidth]{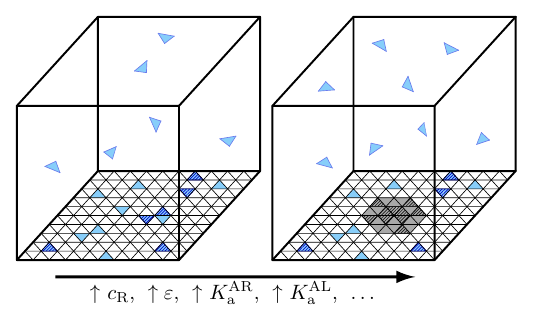}
        \put(3,48){\large \textbf{a}}
    \end{overpic}
    \\
    \begin{minipage}{0.28\linewidth}
        \vspace{0.5em}
        \begin{overpic}[width=\linewidth]{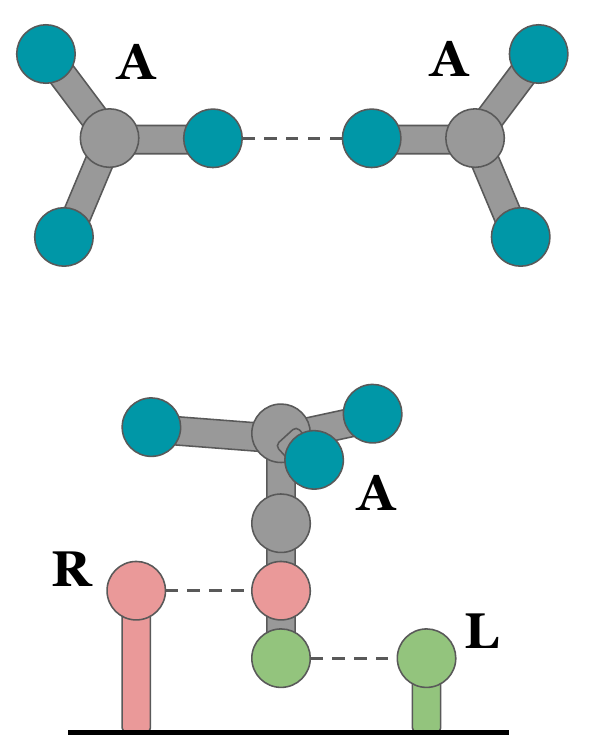}
            \put(17,-5){\footnotesize \textsf{membrane}}
            \put(2,100){\large \textbf{b}}
        \end{overpic}
    \end{minipage}
    \hspace{1em}
    \begin{minipage}{0.6\linewidth}
        \vspace{1.5em}
        \begin{overpic}[width=\linewidth]{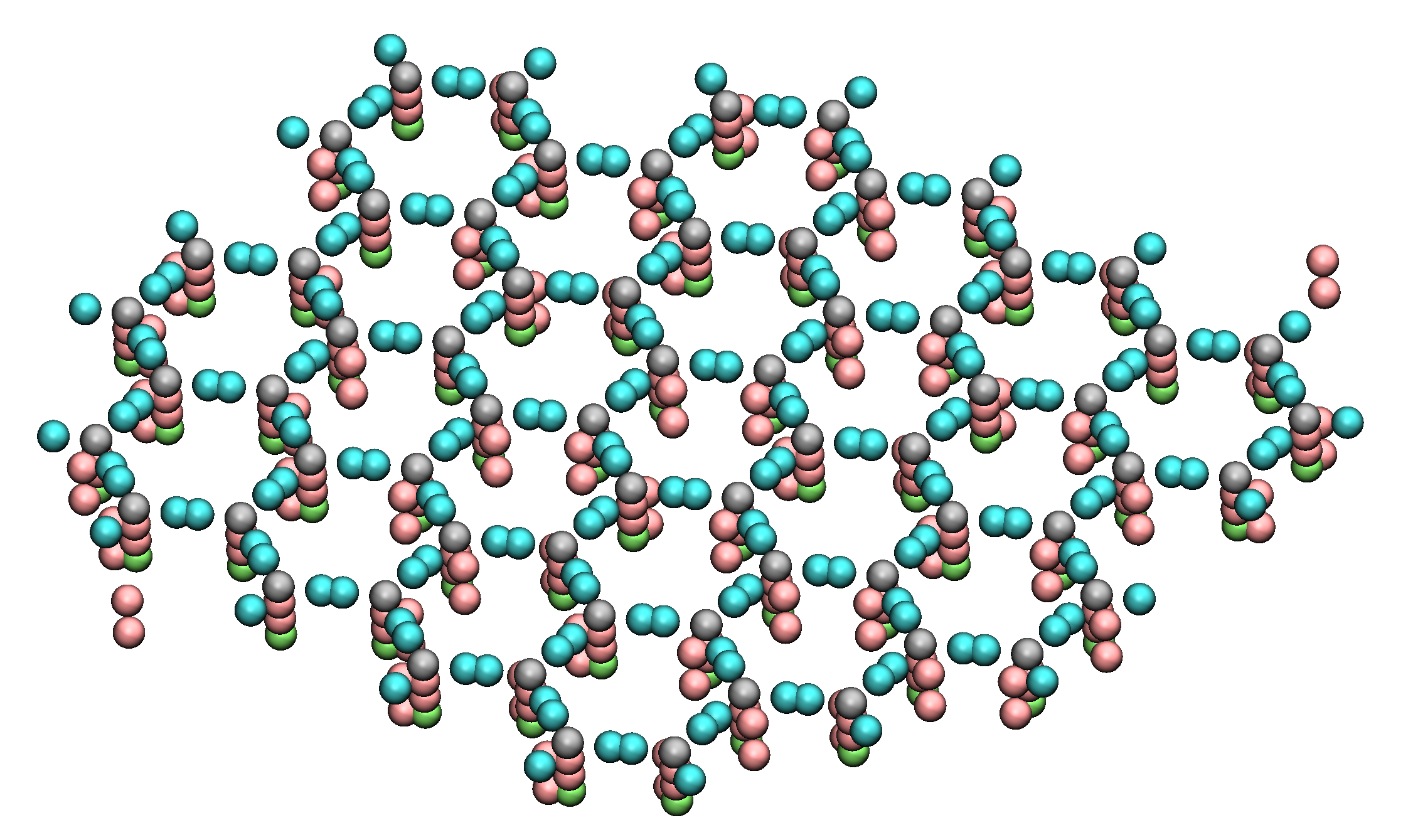}
            \put(70,52){$t=60$ s}
            \put(8,62){\large \textbf{c}}
        \end{overpic}
    \end{minipage}
    \caption{(\textbf{a}) Schematic illustration of lattice theory of receptor-coupled surface self-assembly. The membrane surface area (bottom) is discretized into a lattice ($Z=3$ shown) which may be occupied by A subunits either in the free phase (blue) or lattice phase (grey). Surface subunits bound to a receptor R are indicated with hashed shading.
    (\textbf{b}) Schematic diagram of the coarse-grained model used in reaction-diffusion (RD) simulations.
    (\textbf{c}) Simulation snapshot of an assembled coat 60 seconds after unassembled initiation in solution.
    }
    \label{fig:lattice-figure}
\end{figure}

\section{Models and Methods}

\subsection{Equilibrium Theory}

We construct a minimal model of equilibrium self-assembly coupled to membrane localization and transmembrane receptor binding, deriving our surface free energy from a lattice model. Our assembly-competent subunits A can bind to a single membrane site L 
and may also bind to a transmembrane receptor R once bound to the membrane (Fig 1b) \cite{traub2003sorting}. The subunit valence $Z$ specifies how many contacts subunits can make to other subunits.
The key energetic parameters are the subunit-lipid association constant $\Ka{AL}$, the subunit-receptor association constant $\Ka{AR}$, and the subunit-subunit bond (free) energy $\varepsilon$. The procedure can be summarized as finding the simultaneous solution of the coupled equilibria that govern the state of the assembled structure on the membrane: subunit-membrane equilibrium, subunit-receptor equilibrium, and subunit-subunit surface equilibrium.

The first of these equilibria, for the binding of A and L to form AL, follows simply from the 3-dimensional bimolecular equilibrium constant,
\begin{equation}
	\Ka{AL} = \frac{\cc{AL}}{\cc{A}\cc{L}}.
	\label{eqn:KdAL}
\end{equation}
We treat the concentration $\cc{L}$ as a constant under the reasonable assumption that lipid copies far exceed subunit copies \cite{yogurtcu2018cytosolic}. Throughout this work we will use $\cc{X}\equiv N_\text{X}/\mathcal{V}$ to denote the volume concentration of species X, where $N_\text{X}$ is the copy number. For species which are restricted to move along the 2-dimensional membrane, $\cc{X}$ refers to the effective 3-dimensional concentration, related to the 2-dimensional concentration $\rho_\text{X}\equiv N_\text{X}/\mathcal{A}$ by $\ell$.

The second equilibrium equation is similar, but slightly modified due to the subunit-receptor reaction being confined to the membrane surface. We have\cite{yogurtcu2018cytosolic}
\begin{equation}
	\gamma\Ka{AR} = \frac{\cc{RAL}}{\cc{AL}\cc{R}},
	\label{eqn:KdAR}
\end{equation}
where $\gamma \equiv \ell/h$ is a unitless \emph{dimensionality factor}. The phenomenological dimensional reduction length scale $h$ relates association constants in 2D and 3D, $\Ka{X,3D} = h\,\Ka{X,2D}$\cite{wu2011transforming,xu2015binding}. This parameter accounts for the unavoidable changes to $\Delta G$ in 2D compared to 3D, and is typically on the order of the molecular length scale (\textit{i.e.}, nm). In this work, unless otherwise noted, $\Ka{}$ will refer to the 3D equilibrium constant, being the quantity most often measured experimentally\cite{wu2011transforming}.

To solve the final equilibrium condition of multivalent subunit assembly, we use a discrete 2D lattice for the membrane surface with $M$ total sites of area $a$ (the subunit area). Each site may be occupied by a single subunit, AL or RAL (Fig 1). The total number of free and assembled subunits on the membrane is $N$, and thus $\cc{A,tot}=\cc{A}+N/\mathcal{V}$. We assume a two-state model for the $N$ membrane-bound subunits: the fraction free are defined by $\psi=\mathcal{V}(\cc{AL}+\cc{RAL})/N$, and the fraction in the lattice coat are then given by $1-\psi$. Defining the surface coverage fraction $\phi = N/M = Na/\mathcal{A}$, the approximate subunit entropy $S$, derived in the SI under the assumption of low total subunit coverage $\phi$, is
\begin{equation}
	S = N\kb \left[ \phi \psi^2 + (1-2\phi)\psi - \psi \ln(\phi\psi) \right].
	\label{eqn:entropy}
\end{equation}
We pair this with a simple energy ansatz
\begin{equation}
	E = -\frac{1}{2}(1-\psi)N Z\varepsilon \kb T + E_b.
    \label{eqn:energy}
\end{equation}
The unspecified energy $E_b$ is the energetic penalty of the assembly edge where subunits are not fully saturated with bonds, giving rise to a line tension $\propto \varepsilon \kT / \sqrt{a}$.
Putting together Equations~(\ref{eqn:entropy}) and (\ref{eqn:energy}), our free energy per surface subunit $f = (E-TS)/N\kb T$ is
\begin{equation}
	f = \frac{1}{2}(\psi - 1)Z\varepsilon - \phi\psi^2 + (2\phi - 1)\psi + \psi \ln(\phi\psi) + e_b,
    \label{eqn:free-energy}
\end{equation}
where $e_b = E_b / N\kT$. The equilibrium unassembled fraction $\psi$ then follows from $\partial f / \partial \psi = 0$. The equilibrium value of $\phi$ follows from the coupling to Eqn.~(\ref{eqn:KdAL}), which requires us to re-express the concentrations of surface-bound A species in terms of $\phi$ and $\psi$ (Eqn.~(S2)). 

\subsection{Kinetic Theory}

Receptor density and adhesiveness will also control the timescales of assembly. If the timescale of assembly is too slow for biological relevance, then assembly will not function as a decision-making switch even if it is the thermodynamic equilibrium state. To address this, we use mass-action kinetics to write a system of reaction ODEs modeling the early assembly process (Eqns. S15-S20). This model is parametrized by kinetic rates $\kon{AA}, \koff{AA}$ which are related to our equilibrium constants through the standard relation $\Ka{AA}=\kon{AA}/\koff{AA}$. Association reactions which occur on the 2D membrane surface are effectively accelerated (or decelerated) by the factor $\gamma$ \cite{yogurtcu2018cytosolic}. As the subunit valence $Z$ strongly affects the combinatorial factors appearing in the rate equations, we develop this model specifically for $Z=3$ in order to compare against our simulation model elaborated below.

\subsection{Rigid-Body Model}

To test the reliability of our theoretical predictions under explicit spatial and stochastic fluctuations, we developed a rigid-body model designed for stochastic particle-based reaction-diffusion simulations. The basic elements of the model are illustrated in Fig.~\ref{fig:lattice-figure}b. The multivalent assembly subunit A has three self-binding sites allowing for planar self-assembly into a hexagonal lattice, along with separate binding sites for the lipid L and receptor R molecules. Each of these (L and R) each have only one binding site, being the binding partner of the corresponding site on A. Within our simulations,
we include the one allosteric effect imposed in the theory: subunits bind receptors only after binding to the membrane. Details of the CG geometry are provided in the SI.

\subsection{Stochastic Reaction-Diffusion Simulations}

We performed particle-based reaction-diffusion simulations of the model just described using the NERDSS software\cite{varga2020nerdss}. NERDSS generates stochastic trajectories of rigid-body components consistent with the Smoluchowski model of collision-based reactions\cite{smoluchowski1917mathematical,johnson2014free}. Key inputs for the simulation are rigid molecule geometries with specified binding sites and binding partners, reaction rates, and copy numbers. Binding and unbinding are parameterized by rates $k_{\text{on/off}}$ directly related to equilibrium constants via $\Ka{}=k_{\mathrm{on}}/k_{\mathrm{off}}$.
All inputs to the simulations are thus identical to the theory, excepting the parameter $\varepsilon$, which is determined by an overall best-fit to all simulation data.
For our theory-simulation comparisons, we ran three series of simulations each with a different value of the subunit-lipid affinity $\Ka{AL}$: $0.003\si{\micro\molar}^{-1}$, $0.01\si{\micro\molar}^{-1}$, and $0.033\si{\micro\molar}^{-1}$. For each of these simulation series, we carried out simulations for 24 different amounts of membrane-anchored receptor R, with copy numbers from 15 to 360 (yielding equivalent volume concentrations $\cc{R,tot}$ between $0.025\si{\micro\molar}$ and $0.60\si{\micro\molar}$). In all simulations, the total A and L concentrations are the same at $\cc{A,tot}=0.2\si{\micro\molar}$ and $\cc{L,tot}=30\si{\micro\molar}$ (mimicking appropriate values in live cells; see Discussion). We carried out 4 replica simulations for each parameter combination. Detailed parameters are reported in the SI.

\section{Results}

\begin{figure*}
    \centering
    \begin{minipage}{0.6\linewidth}
        \begin{overpic}[width=\linewidth]{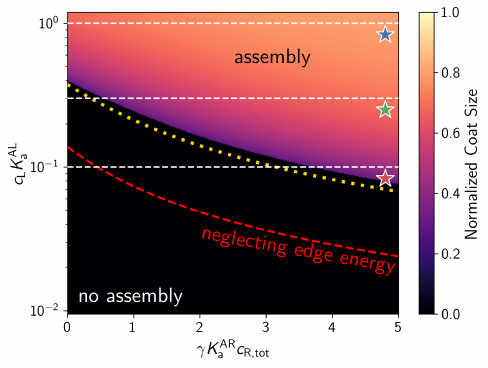}
            \put(2,68){\large \textbf{a}}
        \end{overpic}
    \end{minipage}
    \begin{minipage}{0.255\linewidth}
        \vspace{-0.5em}
        \begin{overpic}[width=\linewidth]{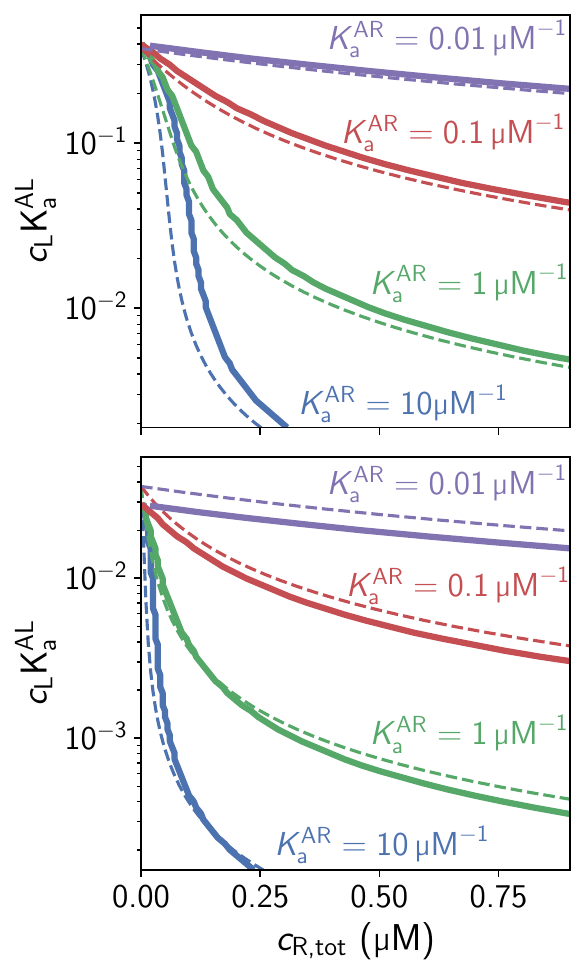}
            \put(2,92){\large \textbf{b}}
        \end{overpic}
    \end{minipage}
    \caption{(\textbf{a}) Assembly phase diagram as adhesiveness (y-axis) and receptor binding (x-axis) are varied. $\cc{A,tot}=\qty{0.2}{\micro\molar}$, $\Ka{AR}=\qty{0.1}{\micro\molar}^{-1}$, $Z=3$, $\varepsilon=5.1$, $h=\qty{10}{\nano\meter}$, and $\ell=\qty{1}{\micro\meter}$. Color shading (lighter is larger) indicates the overall fraction of subunits which have coalesced. The dotted yellow curve being the analytical theory for this boundary given by the rescaled version of eqn.~(\ref{eqn:threshold-r}). The three dashed white lines and colored stars correspond to the series of stochastic simulation results shown in Fig.~{\ref{fig:simulation-details}}a. (\textbf{b}) Top: Assembly thresholds steepen vertically with stronger receptor binding strengths $\Ka{AR}$ as indicated, with other parameters being the same as in the \textbf{a}.
    Bottom: Same as above, but with $\varepsilon=6.8$. Note that the zero-R nucleation point shifts $>$10-fold lower on the adhesion axis. In all cases, exact numerical solutions are solid curves and analytical theory is dashed.}
    \label{fig:phase-diagram}
\end{figure*}

\subsection{Equilibrium Assembly Threshold}

We first examine the predictions of the equilibrium theoretical model numerically. Fig.~\ref{fig:phase-diagram}a plots an assembly ``phase diagram'' showing the numerically-solved assembly size prediction in the adhesiveness-receptor concentration plane. There is a contour (the assembly threshold) along which the assembly size changes discontinuously despite constant $\cc{A,tot}$ and $\varepsilon$. The position and sharpness of the threshold varies with $\Ka{AR}$ and $\varepsilon$ as explored in (Fig.~\ref{fig:phase-diagram}b). 

Our system of coupled equilibria is transcendental, hence why we first approached its solution numerically. However, if one assumes that the edge energy $E_b$ is negligible, it can be solved approximately via series expansion for $\text{exp}(-Z\varepsilon/2)\ll 1$ (see SI), yielding a closed form for the critical receptor concentration $\cc{R}^*$ above which assembly is stable:
\begin{equation}
    \cc{R}^* = \frac{y}{a\ell \xi} \left( \xi - \frac{a h}{y \Ka{AR}} \right) \left( \frac{e^{-Z\varepsilon/2}}{y} - \xi \right),
    \label{eqn:threshold-r}
\end{equation}
in which $\xi = a\ell\cc{A,tot}-\exp(-Z\varepsilon/2)$ and which only depends on $\cc{L}$ and $\Ka{AL}$ through the dimensionless adhesiveness $y\equiv \cc{L}\Ka{AL}$. This threshold is shown as the dashed red curve in Fig.~\ref{fig:phase-diagram}a. Thus the finite-size edge energy $E_b$ cannot be neglected, as it significantly penalizes small lattice sizes, resulting in a change in both the location of the boundary as well as in its character, as the transition is continuous in the absence of the edge penalty. The numerically computed shading in Fig.~\ref{fig:phase-diagram} includes edge energy of the form
\begin{equation}
    E_b = \frac{1}{2}\sqrt{\frac{6 a(1-\psi)}{\mathcal{A}\phi}} \,\varepsilon\kT,
    \label{eqn:boundary-energy}
\end{equation}
corresponding to the lower bound of missing bond energy for $Z=3$ (see SI). The shape of the boundary is largely unchanged, however, and we can rescue equation~(\ref{eqn:threshold-r}) with the \textit{ad hoc} energy rescaling $\varepsilon \rightarrow (1-2/\sqrt{3 \cc{A,tot}\mathcal{V} })\varepsilon$.  This represents an effective average weakening of the binding energy, by taking the edge energy at 50\% assembly and distributing it evenly across all bonds. This modified analytical result produces strong agreement with the exact numerical solution for multiple parameter regimes (Fig.~\ref{fig:phase-diagram}).


\subsection{Bounds and Robustness of Sensing}

Given Eqn.~(\ref{eqn:threshold-r}), we can characterize which systems are capable of sensing receptors and whether the threshold receptor concentration is robust to fluctuations in the membrane adhesiveness. In the absence of R, the coupled equilibria simplify. Then, if the adhesiveness is stronger than a critical strength
\begin{equation}
    y^* =\left(\cc{A,tot} a\ell e^{Z\varepsilon/2}-1\right)^{-1},
    \label{eqn:crit-adh-str}
\end{equation}
then $\cc{R}^*=0$, as no receptors are needed to nucleate assembly. Hence, if $\cc{A,tot}$, $\ell$, or the strength $\varepsilon$ are too high, receptors cannot be sensed. In Fig.~\ref{fig:phase-diagram}a, this corresponds to the region above $\cc{L}\Ka{AL} \approx 0.4$, where assembly is guaranteed regardless of $\cc{R}$. Below this critical strength, the shape of the sensing decision boundary is directly tunable by $\Ka{AR}$ and $h$ via Eqn.~(\ref{eqn:threshold-r}). As subunit-receptor interactions are strengthened by either increasing $\Ka{AR}$ or decreasing $h$, $\cc{R}^*$ decreases. 

The threshold $\cc{R}^*$ is robust to variations in adhesiveness when the decision boundary is sharp, or equivalently when $\mathrm{d}\cc{R}^*/\mathrm{d}y$ is small:
\begin{equation}
    \frac{\mathrm{d}\cc{R}^*}{\mathrm{d} y} = -\left(\frac{e^{-Z\varepsilon/2}}{y^2\gamma \Ka{AR}\xi} + \frac{\xi}{a \ell} \right).
    \label{eqn:threshold-derivative}
\end{equation}
This occurs for strong interactions with receptors (high $\Ka{AR}$ or low $h$), whereas for weak interactions we see the receptor threshold changes rapidly with adhesiveness (Fig.~\ref{fig:phase-diagram}b).  In all cases, when the adhesiveness becomes very weak, the boundary line flattens out as expected, as receptors cannot rescue nucleation. Additionally, either increasing $\ell$ or lowering the solution concentration $\cc{A,tot}$ both sharpen the threshold, supporting detection of low receptor concentrations, albeit also shifting the maximal membrane adhesion strength via Eqn.~(\ref{eqn:crit-adh-str}).

\begin{figure*}
    \centering
    \begin{minipage}{0.5\linewidth}
        \begin{overpic}[width=\linewidth,trim={0 0.5pct 0 0},clip]{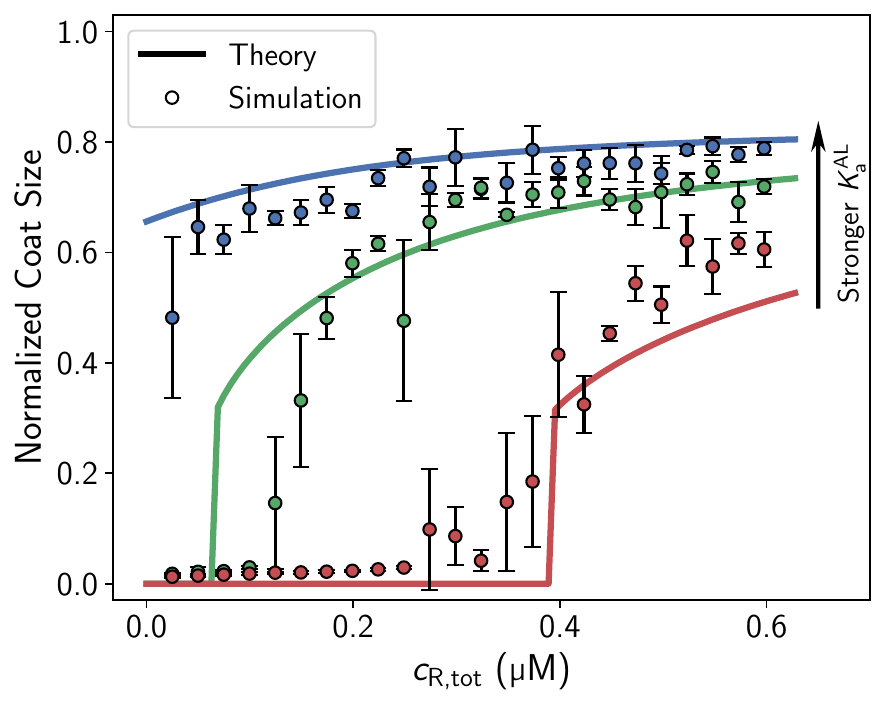}
            \put(2,73){\Large \textbf{a}}
        \end{overpic}
    \end{minipage}
    \begin{minipage}{0.24\linewidth}
        \hfill
        \begin{overpic}[width=0.97\linewidth]{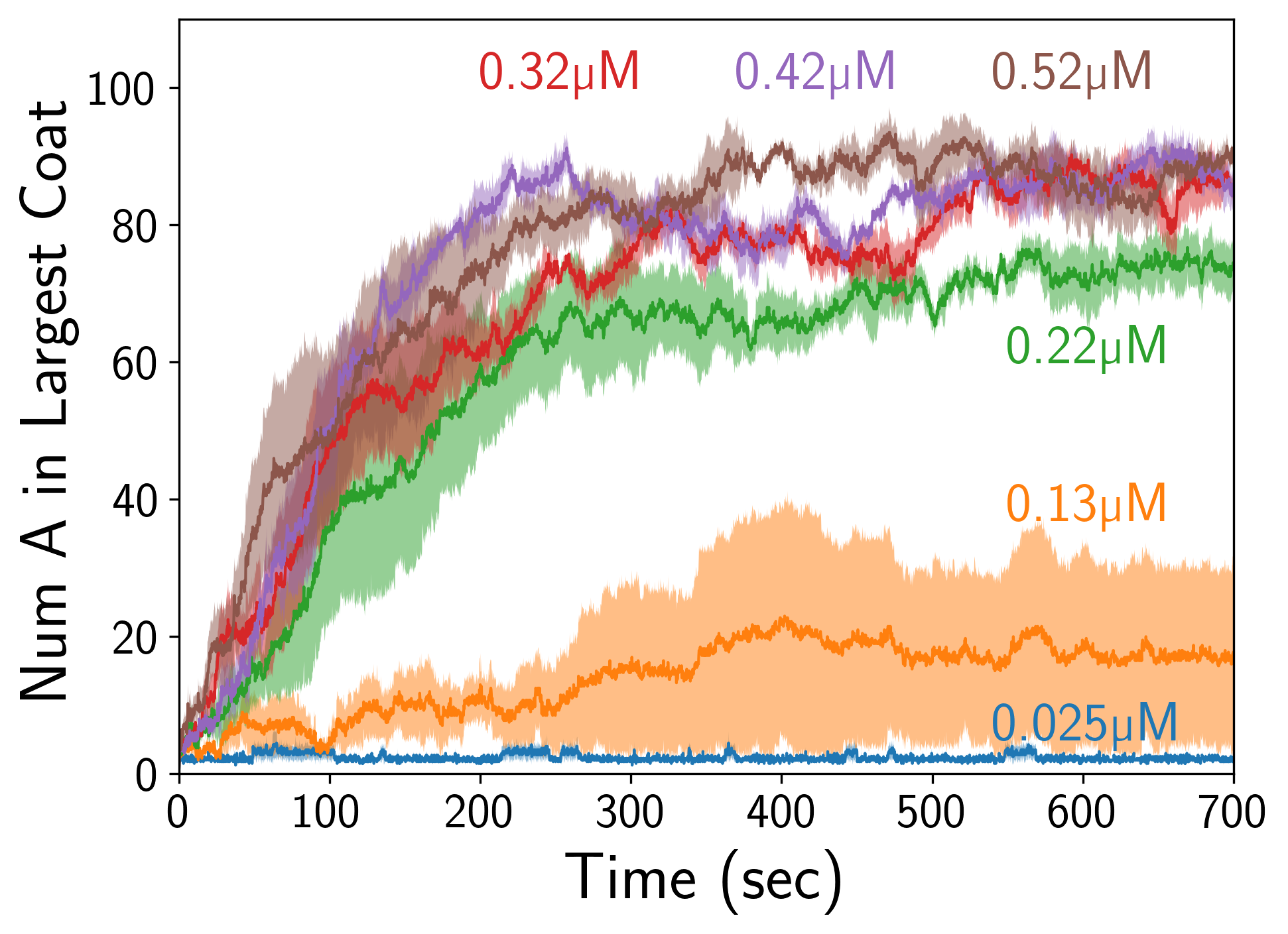}
            \put(-5,64){\Large \textbf{b}}
        \end{overpic}
        \\
        \begin{overpic}[width=0.99\linewidth,trim={0 0.5pct 0 0},clip]{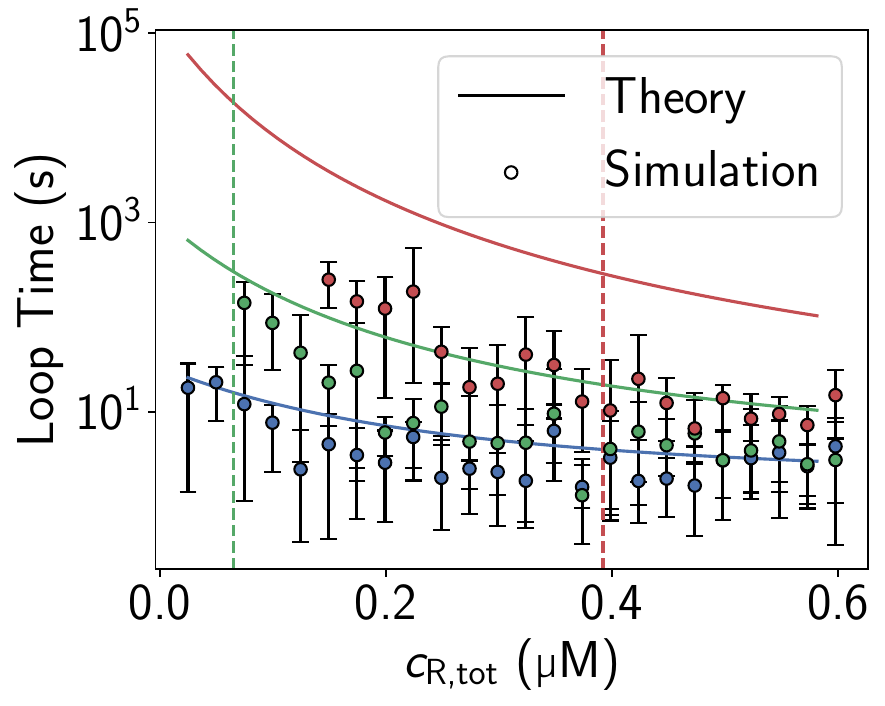}
            \put(-2,68){\Large \textbf{c}}
        \end{overpic}
        \hfill
    \end{minipage}
    \caption{
    RD simulations and theory are in close agreement.
    (\textbf{a}) Equilibrium assembly size as a function of receptor concentration as determined from RD simulations at 24 receptor concentrations. Solid lines are the theoretical result from Eqn.~(\ref{eqn:threshold-r}) with $\varepsilon$-rescaling (see end of Sec. III.A). Points are from RD simulation results. Red, green, and blue correspond to $\Ka{AL}=0.003\si{\micro\molar}^{-1}$, $0.01\si{\micro\molar}^{-1}$, and $0.033\si{\micro\molar}^{-1}$, respectively, for both theory and simulation. Error bars represent the standard error of the mean computed from 4 replica simulations.
    (\textbf{b}) Number of subunits (A) in the largest assembly over time from rigid-body reaction-diffusion simulations with $\Ka{AL}=0.01\si{\micro\molar}^{-1}$. The effective concentration of receptors $\cc{R,tot}$ present in each simulation is annotated for each curve. Results shown are averages from four independent simulations for each curve, with standard error of the mean shown as shading.
    (\textbf{c}) Loop time for hexamer rings measured from RD simulations compared to theory Eqn.~(\ref{eqn:hex-time}). Error bars show the full range of first ring formation times observed in simulation. Dashed lines show transition locations from \textbf{a}. Data points are omitted if any replica never formed a hexamer. Colors are the same as in panel \textbf{a}.
    }
    \label{fig:simulation-details}
\end{figure*}


\subsection{Timescale of Early Assembly Events}

As a proxy for the timescale of initial assembly, we quantify the time $\tau$ to form the first hexagon ring in our $Z=3$ model, the smallest structure with a closed loop of subunits. Our kinetic system (Eqns. S15-S20) is a nonlinear system of 6 ODEs which can readily be solved numerically using standard scientific libraries. However, in order to gain more insight, we use a quasi-steady-state (QSS) approximation to develop an effective 6\ts{th}-order analytical kinetic model, from which we find a timescale of
\begin{equation}
    \tau \approx \left( \frac{729}{8} \mathcal{V} \kon{AA} \gamma^5 {\Ka{AA}}^4 \cc{1,QSS}^6 \right)^{-1},
    \label{eqn:hex-time}
\end{equation}
where $\kon{AA}$ is the rate of A-A binding and $\cc{1,QSS}$ is the approximate QSS concentration of subunits on the membrane, $\cc{AL}+\cc{RAL}$, at early times. The form of this equation resembles results for viral capsids\cite{hagan2014modeling} but with the addition of $\gamma$, and where now $\cc{1,QSS}$ increases with higher adhesiveness and $\Ka{AR}$ to accelerate assembly (see SI sec. 7; Eqn.~(S14)).


\subsection{Stochastic Rigid-Body Reaction-Diffusion Simulations Validate Theory}

We measured the size of the largest assembly throughout the duration of our RD simulations. Example trajectories are shown in Fig.~\ref{fig:simulation-details}b.
From these data, we quantified the size of the largest equilibrium assembly as receptor density increases, with direct comparison to our theoretical model (Fig.~\ref{fig:simulation-details}a). Simulations recapitulate the sharp transition in assembly for finite receptor densities at low adhesiveness, and the persistent nucleation without receptors for high adhesiveness (Fig.~\ref{fig:phase-diagram}b). 
The theory curves in Fig.~\ref{fig:simulation-details}a correspond to the horizontal dashed lines in Fig.~\ref{fig:phase-diagram}a, with the star colors in Fig.~\ref{fig:phase-diagram}a corresponding to the same colors in Fig.~\ref{fig:simulation-details}a. The overall agreement between theory and stochastic simulation is excellent. We found a single best fit value of $\varepsilon$=5.1 through a joint fit to all three data sets in Fig.~\ref{fig:simulation-details}a. We note that this value for $\varepsilon$ agrees well with the 2D binding free energy from simulation, which we estimate in the SI.
We then measured the time to first hexamer ring in each simulation to compare with theory in Fig.~\ref{fig:simulation-details}c. The best agreement is found for stronger membrane binding and low receptor number. For weak membrane binding (red data and curve), Eqn.~(\ref{eqn:hex-time}) over-estimates the time to first hexamer, making the theory a cautious estimator of when assembly timescales will be sufficiently rapid to allow for our assembly sensing mechanism. Importantly, we  note that despite our modest membrane affinity and slow assembly rates ($\kon{AA}=\qty{0.02}{\micro\molar}^{-1}\mathrm{s}^{-1}$), we still see coat initiation within the biologically relevant minutes timescale.


\section{Discussion}

\subsection{Biological and experimental relevance of predictions}
A key prediction of our model is that membrane and receptors are not merely acting as catalysts that lower the barrier to nucleation, but they fundamentally change the stability of the assembled state. As support for this thermodynamic driving, we consider clathrin-mediated endocytosis, an essential process for internalizing receptors that can act as a key step during decision-making within cells\cite{sigismund2021endocytosis}. Our three components L, R and A map to the plasma membrane lipid $\text{PIP}_2$, receptors such as Transferrin Receptor, and the clathrin trimer glued to its adaptor protein. Abundances and affinities used here were previously compiled in refs. \cite{duan2022integrating,guo2022large}. Receptor concentrations can vary widely, but the span we use here (Fig.~\ref{fig:simulation-details}) encompasses Transferrin receptor from distinct cell types (from $\qty{36}{\micro\meter^{-3}}$ to roughly $\qty{1000}{\micro\meter}^{-3}$), and integrin subtypes that reach 10-fold lower still. Because the abundant clathrin trimers ($\sim\qty{0.6}{\micro\molar}$) need adaptors to bind the membrane, we limit the A copies to the abundance of the central adaptor protein AP2 at $\qty{0.2}{\micro\molar}$. The $\Ka{AA}$ value we used is about 4-fold weaker than estimated clathrin-clathrin binding, but this is offset by additional negative cooperativity in clathrin assembly stability needed to reproduce \textit{in vitro} assembly kinetics on membranes\cite{guo2022large}. Finally, $\Ka{AL}$ should capture the strength of adaptor-lipid binding ($\sim \qty{0.1}{\micro\molar}^{-1}$) and clathrin-adaptor binding ($\sim\qty{0.04}{\micro\molar}^{-1}$) for an effective lipid binding strength in the $\Ka{AL}\sim 10^{-3}\,\si{\micro\molar}^{-1}$ range.

Live-cell studies have shown that clathrin assembly requires $\text{PIP}_2$ (L) \cite{boucrot2006role}, and is clearly correlated with receptor levels (R) \cite{mettlen2018regulation}, consistent with our results (Fig.~\ref{fig:simulation-details}). Our model predicts that highly adhesive membranes will trigger clathrin assembly without receptors required (Eqn.~(\ref{eqn:crit-adh-str}) and Fig.~\ref{fig:phase-diagram}). This directly agrees with in vitro experiments; clathrin does not natively assemble in solution\cite{pearse1987structure}, but robustly assembles following adaptor recruitment to lipid membranes with $>5$-fold increases in $\text{PIP}_2$ concentrations compared to the plasma membrane \cite{kelly2014ap2,zeno2021clathrin}. Somewhere between these limits is the phase boundary, and we predict that this transition could be tuned to receptor sensing by driving down $\text{PIP}_2$ levels before adding receptors. Performing this for varying levels of adhesiveness and receptor types would directly probe the shape of Eqn.~(\ref{eqn:threshold-r}).

\subsection{Equilibrium sensing is poor without assembly and dimensional reduction}
\begin{figure}
    \centering
    \includegraphics[width=\linewidth,trim = {0.5em 1em 0.5em 0.5em}, clip]{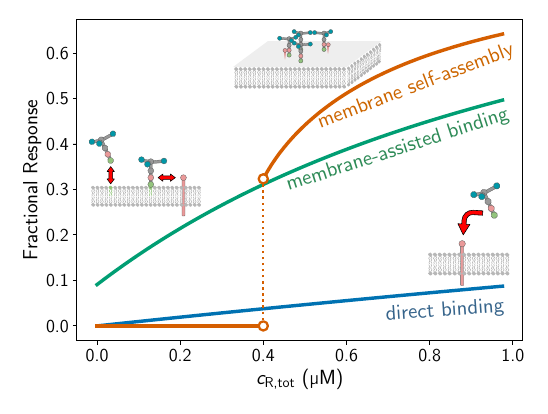}
    \caption{Comparison of different reversible receptor ``sensing'' mechanisms. Direct binding is A binding to R from solution, membrane-assisted binding is A first binding to a lipid L before binding to R. For these two cases, the quantity shown is the fraction of all A molecules which are on the membrane. For the case of membrane self-assembly, the fraction of all A molecules which are in the assembly is plotted. $\Ka{AR}=\qty{0.1}{\micro\molar}^{-1}$ across all examples and $\cc{L}\Ka{AL}=0.1$. }
    \label{fig:sensing-comparison}
\end{figure}

Fig.~\ref{fig:sensing-comparison} compares the sensitivity of self-assembly to that of simpler mechanisms relying on reversible bimolecular interactions, namely direct binding of subunits to target receptors and ``membrane-assisted'' binding, where the subunit first binds to the membrane and then the target receptor, but does not self-assemble. In both cases, the fraction of subunit bound to the membrane (the sensing signal strength) varies with the concentration of the target receptor, with membrane-assisted binding clearly being more sensitive.  That the two-dimensional search is superior to direct binding (reliant on 3D diffusion) is not surprising, as previous work has shown that for our length- and time-scales of interest, two-dimensional diffusion is the energetically optimal mode of information transport\cite{bryant2023physical}. Once 2D assembly is added to the sensing mechanism, the response is both larger in amplitude and exhibits a signal jump at a characteristic threshold, two distinct advantages over the simpler schemes.


\section{Conclusion}

The analytical results presented here neglect fluctuations and inhomogeneities in both A and R, but their good agreement with the explicitly stochastic and spatial simulations that capture diverse assembly sizes (Fig. S4) indicate these assumptions are often broadly valid for predicting sensing and relevant timescales. For situations where receptor distribution is strongly nonuniform, potentially due to self-clustering of receptors, the theory can still be applied on a \emph{local} basis to derive approximate thresholds within specific membrane domains. Such spatial control can be exploited to trigger assembly nucleation at specific locations and with improved cooperativity\cite{hlavacek2001kinetic}, either through direct receptor-receptor interactions\cite{lauffenburger1996receptors}, or through preferential partitioning into a particular membrane phase \cite{rouches2021surface}, e.g., $\mathrm{L_o/L_d}$ phase separation or raft-like domains\cite{levental2016continuing,baumgart2007large}. 
Meso-scale mechanical properties such as membrane rigidity and tension can also couple to the assembly threshold, as the membrane remodeling energy must be provided by the assembling protein machinery\cite{frey2024coat}.

An advantage of membrane-associated self-assembly as a sensing mechanism is to naturally ``package'' the molecules it senses, whether receptors\cite{traub2003sorting} or RNA\cite{cadena2012self}. In this way, the assembly threshold serves a dual purpose also as a \textit{decision threshold}\cite{floyd2025limits}. Moreover, self-assembly is responsive to increases in receptor densities beyond the target threshold (Fig.~\ref{fig:sensing-comparison}). Abrupt increases in external stimuli (such as Calcium) that push well past the assembly threshold would be rapidly (i.e. seconds\cite{thomas2017quantitative}) followed by increasing assembly, as seen in excitable cells\cite{wu2009ca2}, and consistent with the timescales we predict. Maintaining homeostasis in cells, including in cell crawling and endocytosis, requires active energy input to recycle receptors on/off the membrane\cite{mettlen2018regulation}, but our model demonstrates that this necessary activity can be coupled to downhill self-assembly to relax to the homeostatic ``set point'' for receptor density.  The disadvantage of this mechanism is that it is not well-suited for sensitive detection of ultra-dilute components like a single receptor that is possible via signal transduction\cite{bialek2005physical,ten2016fundamental}, as that would not be sufficient to promote assembly. 
Close to the threshold, there are large fluctuations in assembly with slower equilibration times, but living systems may operate near this threshold: For endocytosis, about 50\% of initiated assembly events are observed to be unsuccessful\cite{loerke2009cargo}.

By expanding beyond the three distinct components incorporated here, we expect our model to reproduce the phenomenology of ‘multifarious’ self-assembly models, where diverse subunits facilitate tunable, separable condensates\cite{jacobs2021self}, optimal assembly kinetics\cite{jhaveri2024discovering,jacobs2015rational}, and programmable classification\cite{evans2024pattern}. The same assembly subunit (like clathrin) could be tuned to establish multiple detection limits for distinct receptors by exploiting the sensitivity of nucleation to variations in concentration or affinities, demonstrating nucleation-driven pattern recognition\cite{evans2024pattern,zhong2017associative} in living systems.

\begin{acknowledgments}
The authors thank Markus Deserno for providing feedback on an initial draft of this work. The authors acknowledge support from NIH MIRA Award R35GM133644.
\end{acknowledgments}


\bibliography{apssamp}

\end{document}


\renewcommand{\theequation}{S\arabic{equation}}
\renewcommand{\thefigure}{S\arabic{figure}}

\begin{center}
    {\Large Supplementary Information for: 
    
    Membrane-Associated Self-Assembly for Cellular Decision Making}

    Samuel L. Foley and Margaret E. Johnson
\end{center}

\section{Surface Free Energy}

Consider a 2-dimensional lattice with $M$ total sites that can be occupied by self-assembly monomers A. There are $N$ of these A in total on the surface. We will work in the dilute limit where we assume $N\ll M$. We will denote the number of free individual $A$ monomers as $N_f$, so that $N_f\leq N$. The remaining $N-N_f$ monomers are assumed to reside in a single coat. Here is an example lattice configuration:

\begin{center}
     \includegraphics[width=0.5\linewidth]{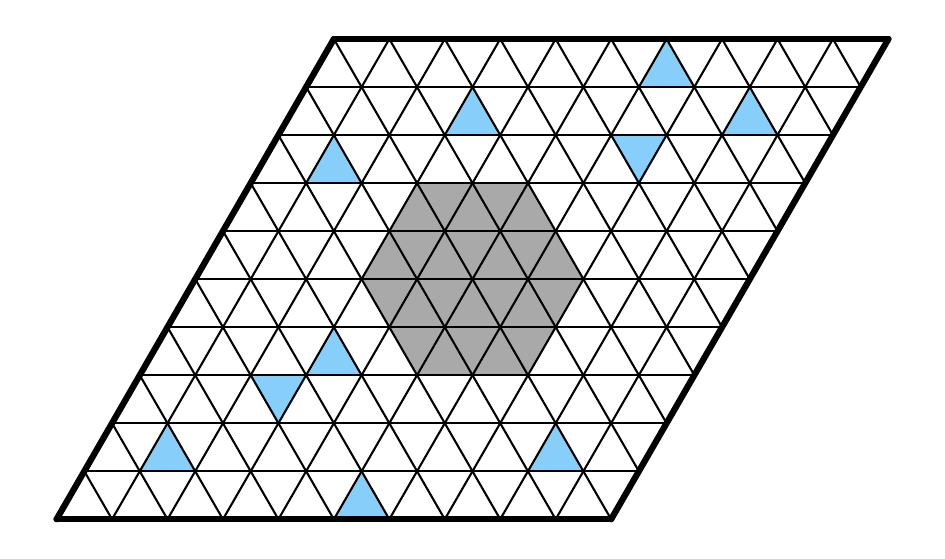}
\end{center}

\noindent In this example, $M=200$, $N=34$, $N_f=10$. The number of ways to arrange the $N_f$ free monomers is
\begin{equation*}
    \binom{M-(N-N_f)}{N_f} = \frac{(M-(N-N_f))!}{N_f!(M-(N-N_f)-N_f)!} = \frac{(M-N+N_f)!}{N_f!(M-N)!}
\end{equation*}
The $O(M)$ positions for the coat represent a small contribution compared to this entropy, so we will ignore the coat entropy entirely. Thus we take as our entropy
\begin{equation*}
    S = \kb \ln\left[ \frac{(M-N+N_f)!}{N_f!(M-N)!} \right].
\end{equation*}
Unpacking this using rules of logarithms and applying Stirling's approximation yields
\begin{equation*}
    \frac{S}{\kb} \approx (M-N+N_f)\ln M + (M-N+N_f)\ln\left( 1 - \frac{N-N_f}{M} \right) - N_f \ln N_f - (M-N)\ln M - (M-N) \ln \left( 1 - \frac{N}{M} \right)
\end{equation*}
Linearizing $\ln (1+x)\approx x$ for small $x$ and simplifying, we have
\begin{equation*}
    \frac{S}{\kb} \approx N_f \ln M + (M-N)\frac{N_f}{M} + N_f\frac{N_f-N}{M} - N_f \ln N_f.
\end{equation*}
We now introduce the following coordinates:
\begin{equation*}
    \phi = \frac{N}{M} \qquad \psi = \frac{N_f}{N}.
\end{equation*}
$\phi$ is the overall surface coverage and $\psi$ is the fraction of free monomers. In the example shown above,  $\phi \approx 0.17$ and $\psi \approx 0.29$. With these, our entropy becomes Eqn.~(3) of the main text:
\begin{equation}
    \frac{S}{N\kb} = \phi \psi^2 + (1-2\phi)\psi - \psi \ln(\phi\psi).
\end{equation}
We make the simplest possible assumption for the energy: each monomer in the coat has $Z$ neighbors ($Z=3$ in our example shown above, as for a clathrin-like assembly), and we will attribute $-\varepsilon \kT$ to each bond, meaning
\begin{equation*}
    E = -\frac{1}{2}(N-N_f)Z\varepsilon\,\kT.
\end{equation*}
Thus, our free energy is
\begin{equation*}
    f = \frac{E-TS}{N\kT} \approx \frac{1}{2}(\psi - 1)Z\varepsilon - \phi\psi^2 + (2\phi - 1)\psi + \psi \ln(\phi\psi).
    \label{eqn:free-energy}
\end{equation*}

\begin{figure}
    \centering
    \includegraphics[width=0.48\linewidth]{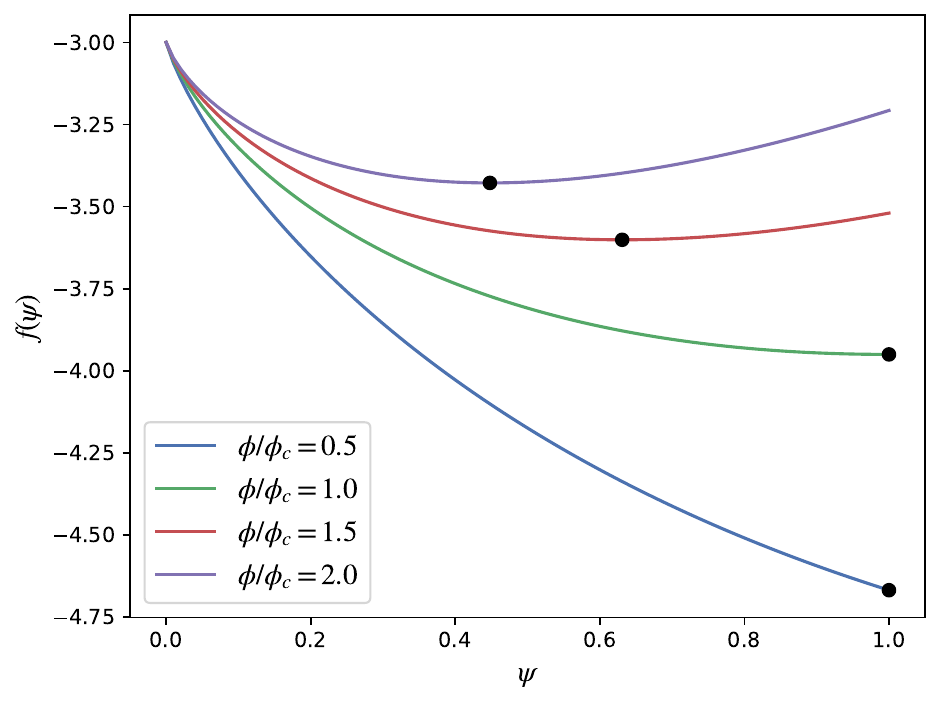}
    \includegraphics[width=0.48\linewidth]{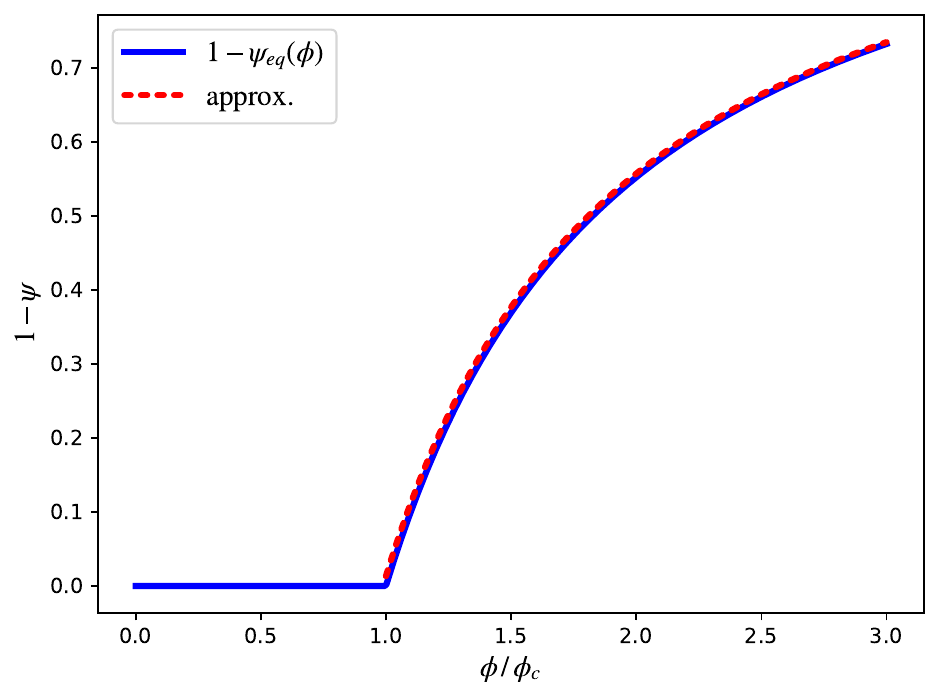}
    \caption{(a) Plots of eqn.~(\ref{eqn:free-energy}) for various values of surface coverage $\phi$ with $\varepsilon = 2$, $Z=3$ ($\phi_c\approx 0.05$). The black point on each curve indicates the free energy minimum. (b) Plot of equilibrium fraction of monomers in the assembled coat, $1-\psi_{eq}$. Dashed red shows the approximation that avoids using special functions.}
    \label{fig:fe-plots}
\end{figure}

Fig.~\ref{fig:fe-plots}a plots $f(\psi)$ for $\varepsilon = 2$, $Z=3$ and a few values of overall surface coverage $\phi$. For this bond energy, coat assembly begins at $\phi_c \approx 0.05$, or 5\% surface coverage, from which point the coat grows with increasing $\phi$. Fig.~{\ref{fig:fe-plots}}b plots the fraction of monomers within the coat, $1-\psi$, as a function of $\phi$, as found by minimizing the free energy. We will now derive analytical expressions for the critical coverage $\phi_c$ at which coat nucleation begins, as well as for the equilibrium fraction of free monomers, $\psi_{eq}$.
\begin{align*}
    \frac{\partial f}{\partial \psi} &= \frac{1}{2}Z\varepsilon - 2\phi\psi + 2\phi + \ln(\phi\psi)
    \label{eqn:d1}
\end{align*}
The equilibrium value of $\psi$ is found by setting this equal to zero, with the restriction that $0\leq \psi \leq 1$. We can find $\phi_c$ by inserting $\psi=1$ and solving for the $\phi$ that makes the above zero,
\begin{equation*}
    \frac{1}{2}Z\varepsilon + \ln(\phi_c) = 0 \quad \rightarrow \quad \phi_c = e^{-Z\varepsilon/2}
\end{equation*}
For $\phi\leq\phi_c$, $\psi_{eq}=1$. For $\phi>\phi_c$, we have to minimize $f$, meaning solving for the roots $\psi(\phi)$ of $\partial f / \partial \psi$, which is transcendental, having solution
\begin{equation*}
    \psi(\phi) = -\frac{1}{2\phi}W_0\!\left( -2e^{-2\phi -Z\varepsilon/2} \right), \qquad \phi > \phi_c \,.
    \label{eqn:psi_lambert}
\end{equation*}
Here $W_k(x)$ is the $k^\text{th}$ branch of the Lambert W function, the inverse of $f(W)=We^W$, with $W_0(x)$ being the ``principal branch.'' This solution was used to plot the blue curve in Fig.~\ref{fig:fe-plots}b. A very good approximation is acquired by series expanding $W_0(x)$ about $x=0$,
\begin{equation*}
    W_0(x) \approx x - x^2
\end{equation*}
\begin{equation*}
    \implies \psi(\phi) \approx \frac{e^{-2\phi-Z\varepsilon/2}}{\phi}\left( 1 + 2e^{-2\phi - Z\varepsilon/2} \right), \qquad \phi > \phi_c.
\end{equation*}
This was used to plot the red dotted curve in Fig.~\ref{fig:fe-plots}b, which is scarcely distinguishable from the exact result. As $\varepsilon$ increases this approximation becomes better. For sufficiently small $\phi_c$ one can also ignore the second term in the expansion: at $\varepsilon=2$ (the case plotted above), it accounts for roughly 9\% of the value, while at $\varepsilon=4$ it has decreased to less than half a percent. For the model simulated in the main text, where $\varepsilon\approx 5.1$, it is entirely negligible. We then take
\begin{equation*}
    \psi_{eq}(\phi) =
    \begin{cases}
        1 & \phi \leq \phi_c
        \\
         \displaystyle\frac{e^{-2\phi-Z\varepsilon/2}}{\phi} &  \phi > \phi_c
    \end{cases}
    \label{eqn:approx_psieq}
\end{equation*}

\section{Membrane Binding Equilibrium}

We will denote the surface area to volume ratio as $\ell = V/A$, and each monomer takes up a surface area $a_0$. The relevant concentration variables in terms of our free energy coordinates $\phi$ and $\psi$ are
\begin{equation*}
    M = \frac{A}{a_0} \implies \conc{A,mem} = \frac{\phi M}{V} = \frac{\phi}{a_0 \ell} \implies \conc{A} = \conc{A,tot} - \frac{\phi}{a_0 \ell} \qquad \conc{AL} = \frac{\phi \psi}{a_0 \ell}, \qquad (\text{No R present})
\end{equation*}
where here $\conc{A,mem}$ is the effective concentration of all A units on the membrane. $\conc{AL}$ is the concentration of the species AL, that is, membrane-bound free monomers. Since it is only these which are free to unbind from the membrane (without having to unbind from other surface-bound partners first), this is the species which enters into the membrane binding equilibrium expression. Solving for $\phi$ from the binding equilibrium given by $\Ka{AL}$ (Eqn. (1) in the main text) gives
\begin{equation*}
    \Ka{AL} = \frac{\conc{AL}}{\conc{A}\conc{L}} = \frac{\phi \psi / a_0 \ell}{\left( \conc{A,tot} - \frac{\phi}{a_0 \ell} \right) \conc{L}} \qquad\rightarrow\qquad a_0 \ell \conc{A,tot} - \phi \left[ 1 + \frac{\psi(\phi)}{\conc{L}\Ka{AL}} \right] = 0
\end{equation*}
This yields the self-consistent value for the equilibrium area coverage $\phi_{eq}$ as a function of total monomer concentration $\conc{A,tot}$ taking into account monomers binding and un-binding from the surface. Taking the approximated form of $\psi_{eq}$, eqn.~(\ref{eqn:approx_psieq}), we can once again solve in terms of the Lambert W function:
\begin{equation*}
    \phi_{eq}^+(\conc{A,tot}) \approx a_0 \ell \conc{A,tot} + \frac{1}{2}W_0\!\left( - \frac{2 \phi_c e^{-2a_0 \ell \conc{A,tot}}}{\conc{L}\Ka{AL}}  \right)
\end{equation*}
Recalling that $\psi_{eq}(\phi)$ is a piecewise function, and that this solution for $\phi_{eq}$ was calculated with the piece that is only valid for $\phi>\phi_c$, the other piece, where $\psi_{eq}=1$, gives
\begin{equation*}
    \phi_{eq}^-(\conc{A,tot}) = \conc{L}\Ka{AL} \frac{a_0\ell\conc{A,tot}}{\displaystyle 1 + \conc{L}\Ka{AL}}.
\end{equation*}
The crossover between the two happens at the point where $\phi_{eq}$ reaches $\phi_c$, which allows us to define the critical total bulk concentration $\conc{A}^*$ required for coat nucleation:
\begin{equation*}
    \conc{A}^* = \frac{\phi_c}{a_0\ell}\left( 1 + \frac{1}{\conc{L}\Ka{AL}} \right) = \frac{e^{-Z\varepsilon/2}}{a_0\ell}\left( 1 + \frac{1}{\conc{L}\Ka{AL}} \right).
\end{equation*}
And with this, we can write
\begin{equation*}
    \phi_{eq}(\conc{A,tot}) = 
    \begin{cases}
        \phi_{eq}^-(\conc{A,tot}) & \conc{A,tot} \leq \conc{A}^*
        \\
        \phi_{eq}^+(\conc{A,tot}) & \conc{A,tot} > \conc{A}^*
    \end{cases}
\end{equation*}
We can also once again write an approximate solution,
\begin{equation*}
    \phi_{eq}^+ \approx a_0 \ell \conc{A,tot} -  \frac{\phi_c e^{-2 a_0 \ell \conc{A,tot}}}{\conc{L}\Kd{AL}} .
\end{equation*}
The fraction of all A monomers which have coalesced is
\begin{equation*}
    \phi M \times (1-\psi) / (\conc{A,tot}\mathcal{V}) =  \frac{\phi(1-\psi)}{a\ell\conc{A,tot}}
\end{equation*}

\section{Receptor Equilibrium}

For this step we will introduce the approximation that the probability for an A to be bound to a receptor R is independent of whether or not it is in the coat. We define the fraction of membrane-bound A which are bound to R as $\alpha$. Then we can say
\begin{equation}
    \conc{RAL} = \frac{\phi\psi\alpha}{a_0\ell} \qquad \text{and} \qquad \conc{AL} = \frac{\phi\psi}{a_0\ell}(1-\alpha).
    \label{eqn:conc-to-greek}
\end{equation}
With this, we can put all of our equilibrium equations into a consistent framework:
\begin{subequations}
    \label{eqn:eq_sys}
    \begin{align}
	\left( \conc{A,tot} - \frac{\phi}{a \ell} \right) \conc{L}\Ka{AL} &= \frac{\phi\psi(1-\alpha)}{a\ell} \label{eqn:KdAL-surf}
    \\
    \left( \conc{R,tot} - \frac{\phi\alpha}{a\ell} \right) \Ka{AR} &= \frac{\alpha}{\gamma\phi(1-\alpha)}
    \label{eqn:KdAR-surf}
    \\
    2\phi\psi - 2\phi - \ln(\phi\psi) &= \frac{1}{2}Z\varepsilon + \frac{\partial e_b}{\partial \psi} \label{eqn:surf_eq}
    \end{align}
\end{subequations}
Eqns.~(\ref{eqn:KdAL-surf}) and (\ref{eqn:KdAR-surf}) correspond to equations~(1) and (2) of the main text, respectively, while Eqn.~(\ref{eqn:surf_eq}) is $\partial f / \partial \psi = 0$. Now we follow the same steps as above. We determine $\alpha$ from $\Ka{AR}$ (Eqn.~(2) in the main text),
\begin{equation*}
    \DF\Ka{AR} = \frac{\frac{\phi\alpha}{a_0\ell}}{\frac{\phi(1-\alpha)}{a_0\ell} \left( \conc{R,tot} - \frac{\phi\alpha}{a_0\ell} \right)} = \frac{\alpha a_0\ell}{(1-\alpha)(a_0\ell\conc{R,tot} - \phi \alpha)}
\end{equation*}
\begin{equation*}
    \implies \alpha(\phi,\conc{R,tot}) = \frac{1}{2\phi} \left[ \phi + a_0\ell\conc{R,tot} + \frac{a_0\ell }{\DF\Ka{AR}} - \sqrt{\left(\phi + a_0\ell\conc{R,tot} + \frac{a_0\ell }{\DF\Ka{AR}}\right)^2 - 4\phi a_0\ell\conc{R,tot}} \; \right]
\end{equation*}
The net result is that now we determine $\phi_{eq}$ by finding the roots of
\begin{equation}
    a_0 \ell \conc{A,tot} - \phi \left[ 1 + \frac{\psi(\phi)}{\conc{L}\Ka{AL}}  (1-\alpha(\phi,\conc{R,tot}))\right] = 0.
    \label{eqn:final-eq-condition}
\end{equation}
From this we can find nucleation boundaries for different parameter combinations. One of the simpler results is for $\conc{A}^*$ considered as a function of total receptor concentration $\conc{R,tot}$, which we can solve for in the same way as before: set $\psi=1$ and $\phi=\phi_c$ and solve,
\begin{equation*}
    \conc{A}^*(\conc{R,tot}) = \frac{\phi_c}{a_0\ell} \left[ 1 + \frac{ 1 - \alpha (\phi_c,\conc{R,tot})}{\conc{L}\Kd{AL}} \right]
\end{equation*}
{\footnotesize
\begin{equation*}
    = \frac{e^{-Z\varepsilon/2}}{a_0\ell} \left[ 1 + \frac{1}{\conc{L}\Ka{AL}}\left( 1 - \frac{e^{Z\varepsilon/2}}{2}a_0\ell \left[ \frac{e^{-Z\varepsilon/2}}{a_0\ell} + \conc{R,tot} + \frac{1}{\DF\Ka{AR}} - \sqrt{\left(\frac{e^{-Z\varepsilon/2}}{a_0\ell} + \conc{R,tot} + \frac{1}{\DF\Ka{AR}}\right)^2 - \frac{4e^{-Z\varepsilon/2} \conc{R,tot}}{a_0\ell}} \; \right] \right) \right].
\end{equation*}
}
The only difference from the no-receptor case is the factor of $(1-\alpha)$ inside the brackets. This will show us how the necessary adaptor amount $\conc{A,tot}$ for coat nucleation changes as we vary the receptor abundance $\conc{R,tot}$.

More interesting is to instead solve for $\conc{R,tot}$ to acquire a direct formula for the receptor-triggered nucleation threshold for a given monomer concentration. The result in this case is
\begin{equation*}
    \conc{R}^*=\frac{\conc{L}\Ka{AL}}{a\ell} \frac{\displaystyle \left(a\ell\conc{A,tot} +\frac{a h}{\conc{L}\Ka{AL}\Ka{AR}} - e^{-Z\varepsilon/2}\right)\left( e^{-Z\varepsilon/2} + \frac{e^{-Z\varepsilon/2}}{\conc{L}\Ka{AL}} - a\ell\conc{A,tot} \right)}{\displaystyle  a\ell\conc{A,tot}-e^{-Z\varepsilon/2}}
\end{equation*}
which is Eqn.~(7) of the main text. To gain quantitative insight into the robustness and sensitivity of this threshold, we can examine its derivtive with respect to the adhesiveness $y \equiv \conc{L}\Ka{AL}$:
\begin{equation}
    \frac{\mathrm{d}\conc{R}^*}{\mathrm{d} y} = \frac{e^{-Z\varepsilon/2}-a\ell\conc{A,tot}}{al}+\frac{e^{-Z\varepsilon/2}}{y^2\DF \Ka{AR}(e^{-Z\varepsilon/2}-a\ell\conc{A,tot})}
    \label{eqn:threshold-derivative}
\end{equation}
Evaluated at the critical adhesiveness given by Eqn. (9) in the main text, this becomes
\begin{equation}
    \frac{\mathrm{d}\conc{R}^*}{\mathrm{d} y} = - \conc{A,tot} + \frac{e^{-Z\varepsilon/2}}{al} - \frac{a h \conc{A,tot}}{\Ka{AR}} e^{+Z\varepsilon/2}+ \frac{1}{\DF \Ka{AR}}
\end{equation}
This derivative is negative, and its magnitude decreases with increasing $\Ka{AR}$ or $\ell$, but increases with increasing $h$ or $\conc{A}$.

\section{Finite-Size Edge Energy}

In our energy, Eqn.~(4) of the main text, we assume a contribution of $-Z\varepsilon$ for every monomer in the assembly. However, those monomers which reside on the edge of the coat will have fewer than $Z$ bonded neighbors. This constitutes a free energy penalty that is proportional to the perimeter. The exact details of this edge free energy penalty depend on the coat geometry, i.e., the form penalty term is $Z$-dependent.

We will treat the $Z=3$ case considered in the main text using the simplest approximation. Let $N_c$ be the number of monomers in the coat. For ``closed shell'' configurations like in the example pictured at the start of this SI ($N_c=6,24,54,96,...$), the number of monomers in the outermost layer is $2\sqrt{6N_c}-6$. The number of these which constitute the ``edge'', meaning they are missing exactly one bond, is $\sqrt{6N_c}$. Thus the penalty to the free energy per monomer on the membrane is
\begin{equation*}
    \Delta f = \frac{\sqrt{6N_c}}{2N}\varepsilon.
\end{equation*}
Extrapolating this to all $N_c$, not just the closed-shell values, gives us a lower-bound approximation for the edge energy. Recognizing that $N_c=N(1-\psi)$ and $N=\mathcal{A}\phi/a$, this becomes
\begin{equation}
    \Delta f = \frac{1}{2}\varepsilon\sqrt{\frac{6 a(1-\psi)}{\mathcal{A}\phi}}.
    \label{eqn:edge-z3}
\end{equation}

\section{Simulation Parameters}

\begin{itemize}
  \item[] Time step: $\Delta t = \qty{1}{\micro\second}$
  \item[] Total iterations: $7.2 \times 10^8 \; \implies \; t_\text{sim} = 720\,\text{s}$
  \item[] Box geometry: $\qty{1}{\micro\meter}\times\qty{1}{\micro\meter}\times\qty{1}{\micro\meter}$
  \item[] Dimensional reduction length scale: $h = \qty{10}{\nano\meter}$
\end{itemize}

\noindent
\textbf{Fixed initial species counts:}
\begin{itemize}
  \item[] $N_\text{L} = 18{,}000$ $\implies \sim 1\%$ lipid L at avg. area per lipid of $0.6\mathrm{nm}^2$
  \item[] $N_\text{A} = 120$ $\implies \conc{A,tot}\approx 0.2\uM$
\end{itemize}

\noindent
\textbf{Simulation series:} 3 simulation series (3 values of $\Kd{AL}$) were performed, each varying the receptor copies:
\begin{itemize}
  \item[] $N_\text{R} = 15$ to $360$ in increments of $15$ (24 values total)
  \item[] Replicas: 4 independent simulations per $N_\text{R}$ value
\end{itemize}

\noindent
\textbf{Rate constants:}

\begin{minipage}{0.33\linewidth}
    \begin{align*}
        \kon{AL} &= 0.3\,\uM^{-1}\sec^{-1} \\
        \koff{AL} &= \Kd{AL} \times \kon{AL} \\
        \Kd{AL} &= \{30,\,100,\,300\}\,\uM
    \end{align*}
\end{minipage}
\begin{minipage}{0.31\linewidth}
    \begin{align*}
        \kon{AR} &= 1\,\uM^{-1}\sec^{-1} \\
        \koff{AR} &= 10\,\sec^{-1}
    \end{align*}
\end{minipage}
\begin{minipage}{0.31\linewidth}
    \begin{align*}
        \kon{AA} &= 0.02\,\uM^{-1}\sec^{-1} \\
        \koff{AA} &= 10\,\sec^{-1}
    \end{align*}
\end{minipage}
\\

\noindent\textbf{Diffusion constants:} Each species has an isotropic translational and rotational diffusion constant:

\begin{minipage}{0.31\linewidth}
    \begin{align*}
        D_\text{A}^\text{trans} &= 25 \;\si{\micro\meter}^2\,\sec^{-1}
        \\
        D_\text{A}^\text{rot} &= 0.5 \; \text{rad}^2\,\si{\micro\second}^{-1}
    \end{align*}
\end{minipage}
\begin{minipage}{0.31\linewidth}
    \begin{align*}
        D_\text{L}^\text{trans} &= 1 \;\si{\micro\meter}^2\,\sec^{-1}
        \\
        D_\text{L}^\text{rot} &= 0
    \end{align*}
\end{minipage}
\begin{minipage}{0.31\linewidth}
    \begin{align*}
        D_\text{R}^\text{trans} &= 1 \;\si{\micro\meter}^2\,\sec^{-1}
        \\
        D_\text{R}^\text{rot} &= 0
    \end{align*}
\end{minipage}
\vspace{1em}

\noindent L and R have no rotational diffusion in simulation because their CG models are rotation-symmetric with respect to their only free axis (only one rotation axis is free due to confinement to the 2D membrane surface).
\\

\noindent\textbf{Simulation CG model geometry (not to scale):}\\
\\
\vspace{1em}
\begin{minipage}{0.45\linewidth}
    {A bound to L and R:\\}
    \vspace{1em}
    \includegraphics[width=\linewidth]{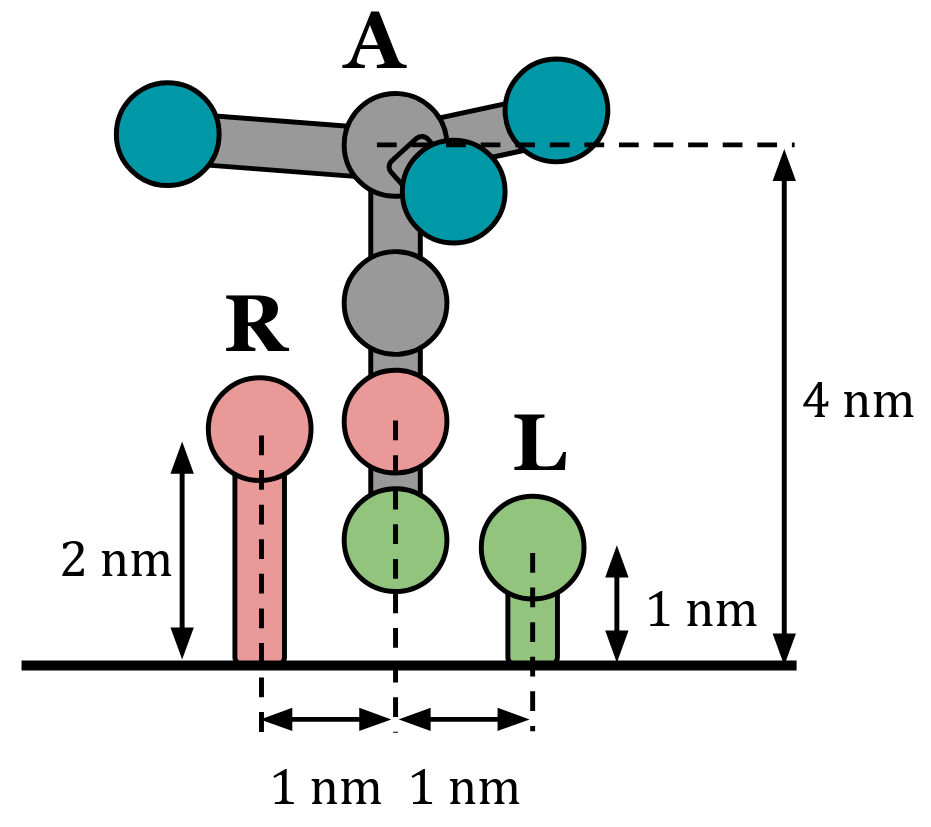}
\end{minipage}
\hfill
\begin{minipage}{0.45\linewidth}
    {\centering Top-down A-A binding: \\}
    \includegraphics[width=\linewidth]{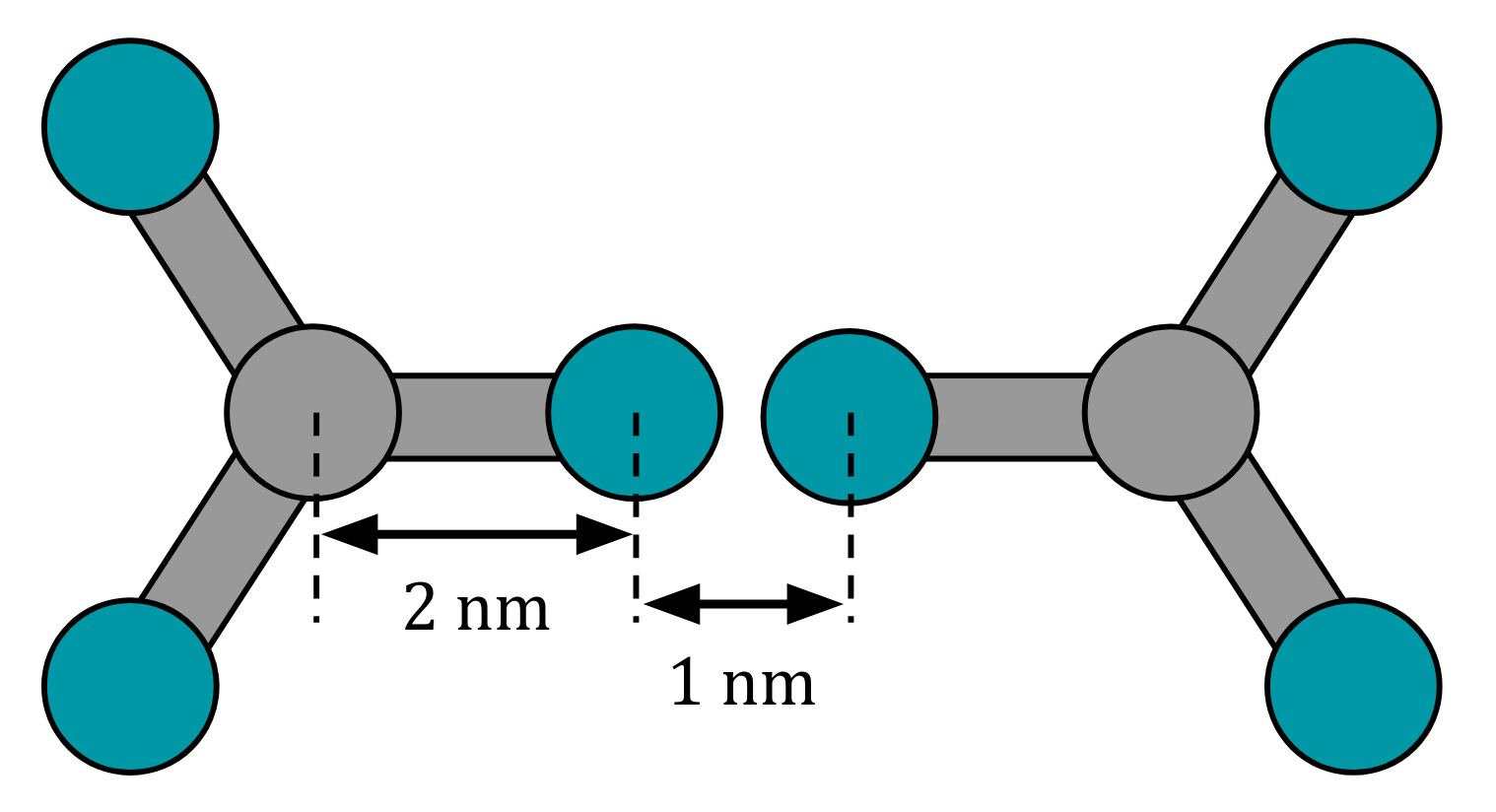}
\end{minipage}

\section{Coat Initiation Time Scale}

\begin{figure}[b]
    \centering
    \includegraphics[width=0.75\linewidth]{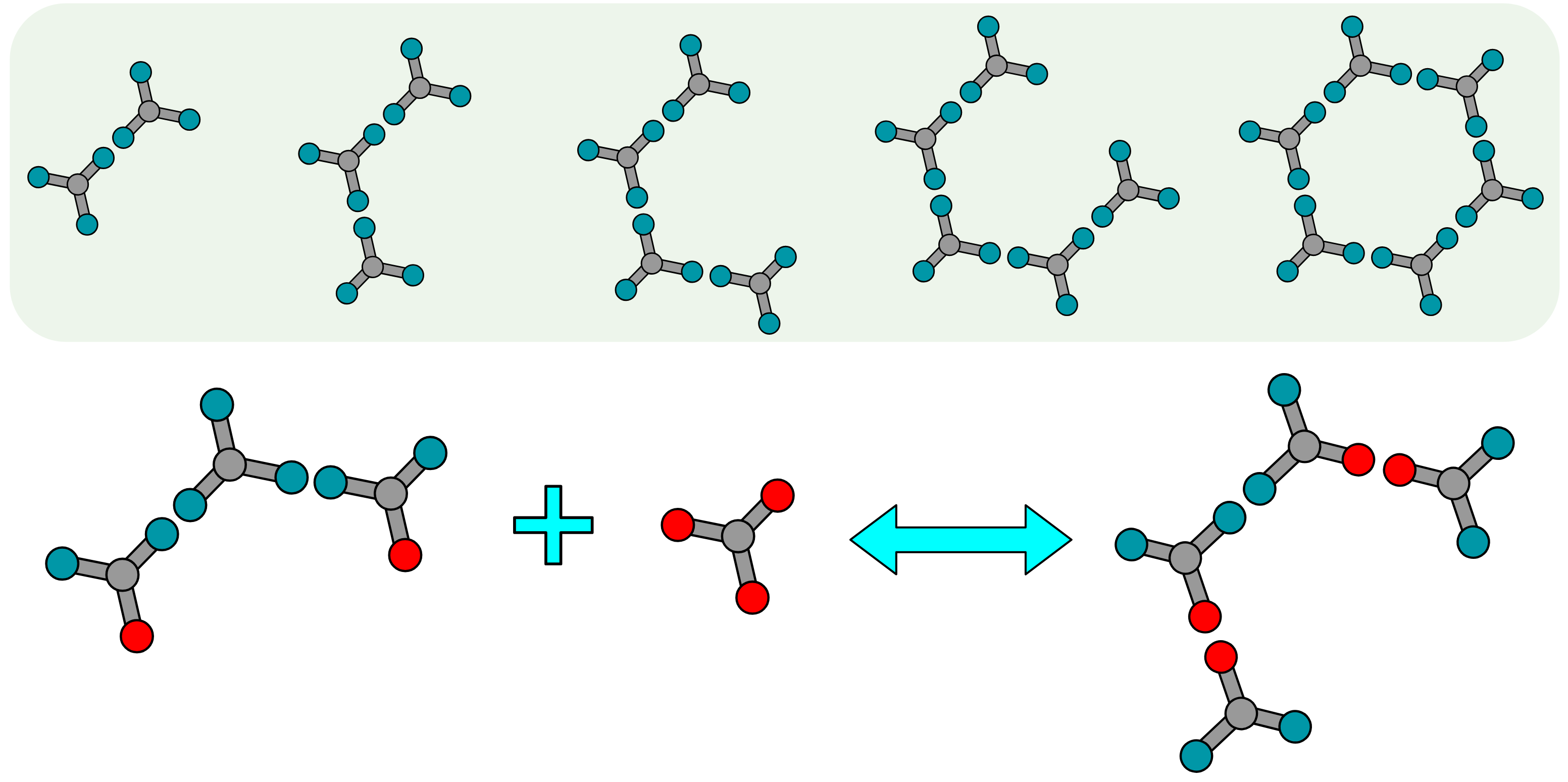}
    \caption{Top: Monomer-addition ring assembly pathway. Bottom: Example step ($\mathrm{A}_3 + \mathrm{A} \leftrightarrow \mathrm{A}_4$) of assembly pathway highlighting in red the structural elements giving rise to combinatorial factors. For the forward reaction, there are $2 \times 3 = 6$ possible associations, and for the reverse reaction there are two bond breakage events that result in loss of a monomer.}
    \label{fig:assembly}
\end{figure}

Here we derive approximate analytical results for the timescale of assembly of a hexagonal A ring on the membrane, the first structure that requires two bond breakages to shrink, making it a natural stable nucleus for coat assembly.

We start by considering the purely 2D single-component problem of the assembly of A monomers into a ring hexamer (Figure \ref{fig:assembly}). Since we are exclusively interested in the time to form a \emph{ring} hexamer, we will only include the most direct assembly pathway in our kinetic calculation (illustrated in the top portion of Figure \ref{fig:assembly}). We work under the assumption of a monomer-addition pathway (all reactions take the form $\mathrm{A}_n + \mathrm{A} \leftrightarrow \mathrm{A}_{n+1}$) in order to derive an approximate expression for the scaling of the nucleation time, to which we add an empirical factor to bring the prediction into quantitative agreement with numerical ODE results which include non-monomer assembly steps. To begin our analytical development, we use a quasi-steady-state (QSS) approximation where $\mathrm{A}_1$ through $\mathrm{A}_5$ have mutually equilibrated. Then we have:
\begin{equation}
    \frac{\conc{2}}{\conc{1}^2} = \frac{9 \gamma\kon{AA}}{\koff{AA}} = 9 \gamma\Ka{AA}, \qquad \frac{\conc{3}}{\conc{2}\conc{1}} = \frac{12 \gamma\kon{AA}}{2 \koff{AA}} = 6 \gamma\Ka{AA}, \qquad \frac{\conc{4}}{\conc{3}\conc{1}} = \frac{6 \gamma\kon{AA}}{2 \koff{AA}} = 3 \gamma\Ka{AA}, \qquad \frac{\conc{5}}{\conc{4}\conc{1}} = 3 \gamma\Ka{AA}
    \label{eqn:qss}
\end{equation}
\begin{equation*}
    \implies \conc{2} = 9\gamma\Ka{}\conc{1}^2, \qquad \conc{3} = 54 \gamma^2 \Ka{2} \conc{1}^3, \qquad \conc{4} = 162 \gamma^3 \Ka{3} \conc{1}^4, \qquad \conc{5} = 486 \gamma^4 \Ka{4} \conc{1}^5
\end{equation*}
In this and all that follows, $\Ka{}$ without a label is assumed to be the A-A binding site affinity $\Ka{AA}$. We may then solve for $\conc{1}$ by substituting the above expressions for $\conc{2}$ through $\conc{5}$ into the mass conservation equation:
\begin{equation}
    \conc{1} + 2\conc{2} + 3\conc{3} + 4\conc{4} + 5\conc{5} = \conc{1} + 18 \gamma\Ka{}\conc{1}^2 + 162 \gamma^2 \Ka{2} \conc{1}^3 + 648 \gamma^3 \Ka{3} \conc{1}^4 + 2430 \gamma^4 \Ka{4} \conc{1}^5 = \conc{tot}^{\mathrm{2D}},
    \label{eqn:mass}
\end{equation}
where $\conc{tot}^{\mathrm{2D}}$ is the total monomer concentration (in the 2D system). Equation (\ref{eqn:mass}) is however a degree 5 polynomial  which therefore has no closed-form analytical solution. We do however expect $\Ka{}\conc{1} \ll 1$ in our parameter regime, and therefore we truncate the polynomial at quadratic order to derive the approximate result
\begin{equation}
    \conc{1} \approx \frac{1}{2} \left( -\frac{1}{18\gamma\Ka{}} + \sqrt{\left( \frac{1}{18\gamma\Ka{}} \right)^2 + \frac{2\conc{tot}^{\mathrm{2D}}}{9\Ka{}} } \right).
\end{equation}
Neglecting the slower backward reaction, we then take the rate of formation of hexamer rings to be
\begin{equation}
    \frac{\mathrm{d}\conc{6}}{\mathrm{d}t} \approx 6 \gamma \kon{AA} \conc{5} \conc{1} \approx 2916 \kon{AA} \gamma^5 \Ka{4} \conc{1}^6 \approx \frac{729}{16} \kon{AA} \gamma^5 \Ka{4} \left( -\frac{1}{18\gamma\Ka{}} + \sqrt{\left( \frac{1}{18\gamma\Ka{}} \right)^2 + \frac{2\conc{tot}^{\mathrm{2D}}}{9\Ka{}} } \right)^6.
    \label{eqn:rate-result}
\end{equation}

Now, we determine the appropriate value for $\conc{tot}^{\mathrm{2D}}$. Neglecting for the moment A+A and A+R binding, the solution-membrane equilibrium for A is determined by the reaction
\begin{equation}
    \mathrm{A} + \mathrm{L} \longleftrightarrow \mathrm{AL}
    \label{rxn:AL}
\end{equation}
with forward rate $\kon{AL}$ and reverse rate $\koff{AL}$. In all cases of interest here, $\conc{L}\gg \conc{A}$, and so we may appeal to the pseudo-unimolecular limit when approximating the relaxation time for this reaction\cite{mishra2021speed},
\begin{equation}
    \tau_\mathrm{AL} \approx \frac{1}{\koff{AL} + \kon{AL}\conc{L}}.
\end{equation}
For our parameter regime, $\tau_\mathrm{AL} < 0.1$ sec. As this is a negligible fraction of the ring nucleation times observed in simulation, we will take the total amount of membrane-bound A as being its equilibrium value as determined by reaction (\ref{rxn:AL}) coupled to receptor binding,
\begin{equation*}
    \mathrm{AL} + \mathrm{R} \longleftrightarrow \mathrm{RAL}
\end{equation*}
with forward rate $\kon{AR}$ and reverse rate $\koff{AR}$. This amounts to solving the system given by equations (1) and (2) of the main text combined with the mass conservation conditions
\begin{equation*}
    \conc{A} + \conc{AL} + \conc{RAL} = \conc{A,tot},
\end{equation*}
\begin{equation*}
    \conc{R} + \conc{RAL} = \conc{R,tot}.
\end{equation*}
The value for $\conc{AL}$ derived from this system is then inserted into equation (\ref{eqn:rate-result}) as $\conc{tot}^{\mathrm{2D}}$ to determine the early time steady-state rate of formation of hexamer rings (see equation (\ref{eqn:analytical-approx}) below). The approximate time scale for the formation of one ring is then $\tau = 1/J$ where
\begin{equation*}
    J = \mathcal{V} \frac{\mathrm{d}\conc{6}}{\mathrm{d}t}.
\end{equation*}

\begin{figure}
    \centering
    \includegraphics[width=0.5\linewidth]{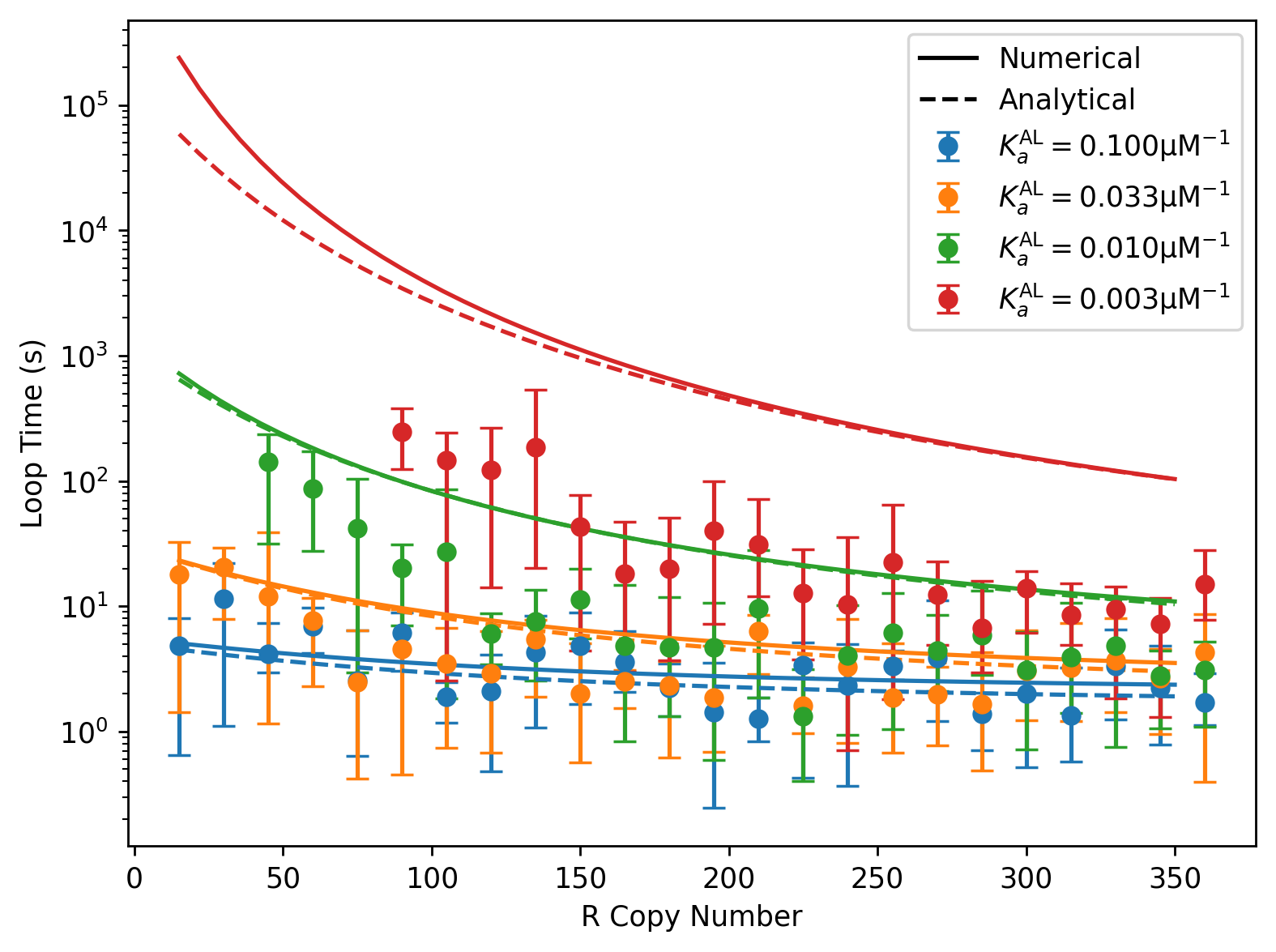}
    \caption{Comparison of hexamer ring formation time measured from NERDSS simulations compared to adjusted theory. Data points represent the mean first ring time in simulation, error bars show the full range of first ring formation times observed in simulation. Data points are omitted if one or more replicas never formed a ring hexamer.}
    \label{fig:looptimes}
\end{figure}

Due to the number of approximations present, we expect this result to differ quantitatively from the simulation results, while still capturing appropriate scaling and qualitative behavior. We therefore compare our analytical approximation just derived with a similar calculation carried out by directly numerically solving a system of nonlinear reaction ODEs for the 2D ring assembly process, thereby eliminating the QSS and monomer-addition approximations. The explicit form of the ODE system is given at the end of this section. We find empirically that a factor of 2 increase in the analytical rate prediction gives good agreement between the analytical and numerical theory results, as shown in Figure \ref{fig:looptimes}. That is, the analytical prediction plotted in the figure, and which appears as Eqn.~(9) of the main text, is
\begin{equation}
    \tau = \left( \frac{729}{8} \mathcal{V} \kon{AA} \gamma^5 \Ka{4} \left( -\frac{1}{18\gamma\Ka{}} + \sqrt{\left( \frac{1}{18\gamma\Ka{}} \right)^2 + \frac{2\conc{tot}^{\mathrm{2D}}}{9\Ka{}} } \right)^6 \right)^{-1},
    \label{eqn:analytical-approx}
\end{equation}
\begin{equation*}
    \conc{tot}^{\mathrm{2D}} = \frac{1}{2}\frac{ \conc{A,tot} - \conc{R,tot} - \Kd{eff} + \sqrt{\left(\conc{R,tot}-\conc{A,tot}+\Kd{eff} \right)^2 + 4\Kd{eff}\conc{A,tot}} }{1 + (\Ka{AL}\conc{L})^{-1}},
\end{equation*}
\begin{equation*}
    \Kd{eff} = \frac{1}{\gamma \Ka{AR}} \left( 1 + \frac{1}{\Ka{AL}\conc{L}} \right).
\end{equation*}

As can be seen from Figure \ref{fig:assembly} and Figure 3b of the main text, the theory compares well to the simulation data for fast assembly times and low R copy number. The discrepancy for larger R copy number can be understood due to the fact that the A-R equilibrium is not directly coupled to the assembly kinetics in the analytical approximation, as this would become analytically intractable. Our analytical expression serves well as an upper bound on first ring assembly time, allowing for an approximate check as to whether the equilibrium assembly `decision' of the main text is suppressed by slow kinetics.

\subsection{Ring Assembly ODE System}

In the following system, all reactions occur on the 2D membrane surface, and as such the factor of $\gamma$ is absorbed into $\kon{}$ for brevity. The reaction system includes ring formation via $\mathrm{A}_2 + \mathrm{A}_4$ and $\mathrm{A}_3 + \mathrm{A}_3$ pathways. Hexamerization is treated as irreversible on the timescale of interest. All stoichiometric and combinatorial factors are included in the coefficients.

\begin{align}
\diff{\conc{1}}{t}
&=
k_{\mathrm{off}}\left(
2 \conc{2}
+ 2 \conc{3}
+ 2 \conc{4}
+ 2 \conc{5}
\right)
-
k_{\mathrm{on}} \conc{1}\left(
18 \conc{1}
+ 12 \conc{2}
+ 6 \conc{3}
+ 6 \conc{4}
+ 6 \conc{5}
\right),
\\[1em]
\diff{\conc{2}}{t}
&=
- 12 k_{\mathrm{on}} \conc{2} \conc{1}
+ 2 k_{\mathrm{off}} \conc{3}
+ 9 k_{\mathrm{on}} \conc{1}^2
- k_{\mathrm{off}} \conc{2}
- 4 k_{\mathrm{on}} \conc{2} \conc{4}
+ 2 k_{\mathrm{off}} \conc{4}
+ 2 k_{\mathrm{off}} \conc{5},
\\[1em]
\diff{\conc{3}}{t}
&=
- 6 k_{\mathrm{on}} \conc{3} \conc{1}
+ 2 k_{\mathrm{off}} \conc{4}
+ 12 k_{\mathrm{on}} \conc{2} \conc{1}
- 2 k_{\mathrm{off}} \conc{3}
- 4 k_{\mathrm{on}} \conc{3}^2
+ 2 k_{\mathrm{off}} \conc{5},
\\[1em]
\diff{\conc{4}}{t}
&=
- 6 k_{\mathrm{on}} \conc{4} \conc{1}
+ 2 k_{\mathrm{off}} \conc{5}
+ 6 k_{\mathrm{on}} \conc{3} \conc{1}
- 3 k_{\mathrm{off}} \conc{4}
- 4 k_{\mathrm{on}} \conc{2} \conc{4},
\\[1em]
\diff{\conc{5}}{t}
&=
- 6 k_{\mathrm{on}} \conc{5} \conc{1}
+ 6 k_{\mathrm{on}} \conc{4} \conc{1}
- 4 k_{\mathrm{off}} \conc{5},
\\[1em]
\diff{\conc{6}}{t}
&=
6 k_{\mathrm{on}} \conc{5} \conc{1}
+ 4 k_{\mathrm{on}} \conc{2} \conc{4}
+ 2 k_{\mathrm{on}} \conc{3}^2.
\end{align}

\section{Heterogeneity of Assembly}

\begin{figure}[H]
    \centering
    \includegraphics[width=0.95\linewidth]{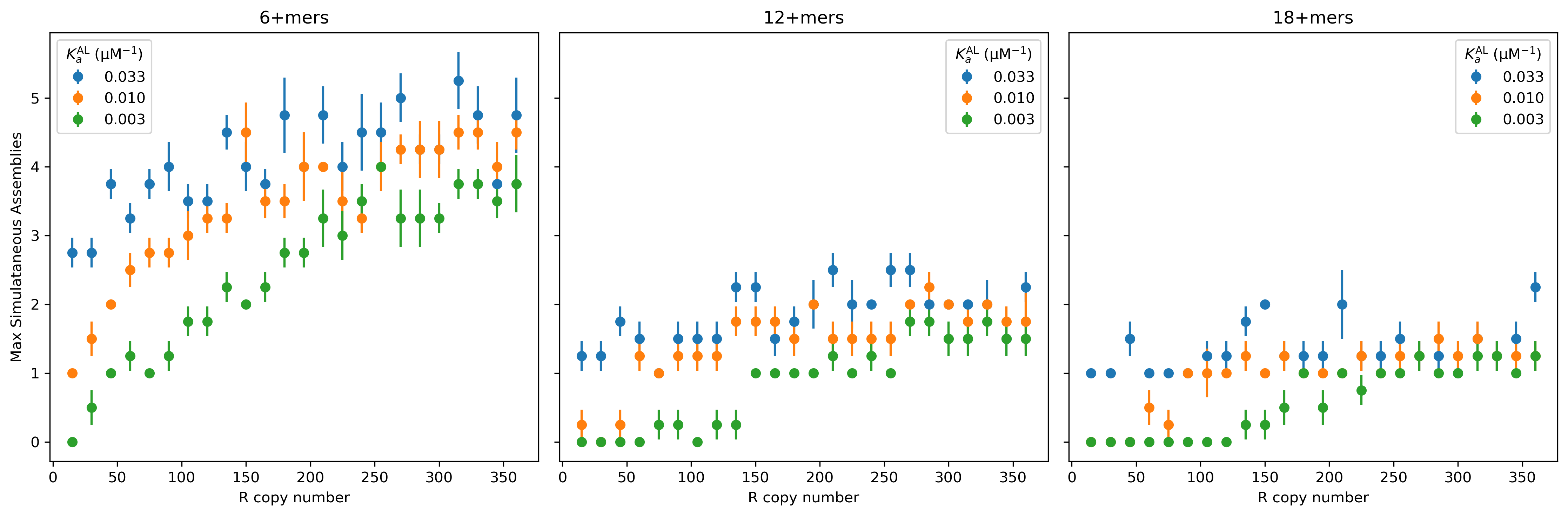}
    \caption{Maximum number of simultaneously coexisting assemblies of the indicated sizes during simulation. The left plot shows the highest number of coexisting assemblies containing 6 or more monomers, the middle plot 12 or more monomers, the right plot 18 or more monomers. An intuitive trend is easy to see, wherein the more favorable it is to assemble a coat, the more simultaneous coats that form. However, due to simulation conditions, most larger coats coalesce into a small number of coats.}
    \label{fig:multi-assembly}
\end{figure}

\section{Estimating a 2D free energy for $\varepsilon$}

The one parameter in the theory that does not immediately map to a parameter in the simulated reaction-diffusion model is the monomer-monomer interaction energy $\varepsilon$. This parameter was therefore used as a fit parameter in optimizing theory against the equilibrium simulation data. To speculate about this interaction strength, we note the following three thermodynamic relationships
\begin{equation}
\Kd{3D}=\frac{1}{\mathcal{V}_{\text{ref}}} e^{\Delta G^{\text{3D}}(\mathcal{V}_{\text{ref}})/\kT}
\end{equation}

\begin{equation}
\Kd{2D} = h \Kd{3D}
\label{eqn:h2D}
\end{equation}
\begin{equation}
\Kd{2D}=\frac{1}{\mathcal{A}_{\text{ref}}} e^{\Delta G^{\text{2D}}(\mathcal{A}_{\text{ref}})/\kT}
\end{equation}

From simulations, we know the input values $\Kd{AA,3D}=\qty{500}{\micro\molar}$ and $h=\qty{0.01}{\micro \meter}$. This therefore uniquely defines $\Kd{AA,2D}$ using Eqn.~(\ref{eqn:h2D}). However, to define $\Delta G^{\text{2D}}$ we must specify the reference area, and this is the problem. A reference area is not defined for the theoretical energy scale $\varepsilon$.

The common reference state in 3D used to define the standard state binding free energy $\Delta G^\circ$ is
$c_0 = \qty{1}{\molar} = 2/\mathcal{V}_\mathrm{ref}$ (2 monomers),
which in this case gives 
$\Delta G^\text{3D} = -6.9\kT$.
If we assume this 3D reference volume defines a standard state length scale $\mathcal{V}_\text{ref}^{1/3}$, then for the reference area we have 
$\mathcal{A}_\text{ref}=2.23\times 10^{-6}\si{\micro\meter}^{2}$,
yielding 
$\Delta G^\text{2D}(\mathcal{A}_\text{ref})\approx -5\kT$.
The dimensionless interaction energy $\varepsilon$ would then be approximately 5. To summarize, this approach is limited by the implied correspondence between an interaction energy and a binding free energy at a specific reference area, here of size $2.23\si{\nano\meter}^2$. These arguments indicate that the free energy between theory and simulation are comparable.

\printbibliography